\documentclass[journal,11pt,onecolumn,draftclsnofoot,]{article}
\usepackage{cite}
\usepackage{amsmath,amssymb,amsfonts}
\usepackage{algorithmic}
\usepackage{graphicx}
\usepackage{textcomp}
\usepackage{xcolor}
\usepackage{tikz}
\usepackage{overpic}
\usepackage{lscape}
\usepackage{rotating}
\usepackage[caption=false, font=footnotesize]{subfig}

\usepackage{geometry}
\usepackage{setspace}   
\usepackage{adjustbox}
\usepackage{hyperref}
\usepackage{xcolor}
\usepackage{siunitx}
\usepackage{xparse}%
\usepackage{cite}
\usepackage{authblk}
\usepackage{multicol}
\usepackage{booktabs}
\usepackage{xcolor,colortbl}
\usepackage[T1]{fontenc}
\usepackage{lipsum}
\usepackage[utf8]{inputenc}
\usepackage{array}
\usepackage{lettrine}
\usepackage{booktabs}
\usepackage{authblk}
\usepackage{graphicx}
\usepackage{stfloats}
\usepackage{amsmath}
\usepackage{graphicx}
\usepackage{amssymb}
\usepackage[round, sort, numbers]{natbib}
\usepackage[mathletters]{ucs}
\usepackage{algorithm}
\usepackage{tabularx,booktabs}
\newcolumntype{C}{>{\centering\arraybackslash}X} 
\usepackage{lipsum}
\newcolumntype{P}[1]{>{\centering\arraybackslash}p{#1}}
\newcommand{\dd}{\mathrm{d}}
\def\BibTeX{{\rm B\kern-.05em{\sc i\kern-.025em b}\kern-.08em
    T\kern-.1667em\lower.7ex\hbox{E}\kern-.125emX}}
\begin{document}

\title{Signal denoising  based on the
Schr\"{o}dinger  operator's eigenspectrum  and a curvature constraint\\
}
\author{{ Peihao Li and Taous-Meriem Laleg-Kirati}
 \affil{The authors are with  Computer, Electrical and Mathematical Sciences and Engineering Division, King Abdullah university of science and technology, KSA. Email: {\tt\small taousmeriem.laleg@kaust.edu.sa}}}
 \date{}
\maketitle
\doublespacing

\begin{abstract} Recently, a new Signal processing method, named Semi-Classical Signal Analysis (SCSA), has been proposed for denoising Magnetic Resonance Spectroscopy (MRS) signals. It is based on the Schr\"{o}dinger Operator's eigenspectrum. It allows an efficient noise reduction while preserving MRS signal's peaks. 
In this paper, we propose to extend this approach  to different signals, in particular pulse shaped signals,  by including an optimization that considers  curvature constraints. The performance of the method is measured by analyzing noisy signal data and comparing with other denoising methods. Results indicate that the proposed method not only produces good denoising performance but also guarantees the peaks are well preserved in the denoising process.
\end{abstract}


\section{Introduction}
The  Semi-Classical Signal Analysis (SCSA) method decomposes the signal into a set of functions given by the squared eigenfunctions of  the Schr\"{o}dinger operator  associated to its negative eigenvalues \cite{b2}. Thus,  and  unlike traditional signal decomposition tools, the  SCSA expresses the signal through a set of functions that are signal dependent, i.e  these functions are not fixed and known in advance  but are computed by solving the spectral problem of the  Schr\"{o}dinger operator whose potential is  the signal to be analyzed. These eigenfunctions will capture more details about the signal and  its morphological  variations. 

\noindent The SCSA has been successfully applied in many  applications for signal representation, denoising,  post-processing and feature extraction. For example, it has been used  for arterial blood pressure waveform analysis in \cite{b2}, \cite{b3}, \cite{b4} and  for Magnetic Resonance Spectroscopy (MRS) denoising \cite{b1} and MRS water suppression \cite{b5}. Is has been also used for feature extraction in epileptic seizure detection using Magnetoencephalography (MEG) signals \cite{b6}.

\noindent  The SCSA has shown very good performance in analyzing pulse-shaped signals that can be found in many applications,  in particular, in biomedical applications. For this type of signals, the peak shape as well as the peak position are of paramount importance \cite{b7}. However, data   come with noise and error, with a variety of noise origins such as electronic, mechanical and optical interferences, causing signal spectrum to be noisy regardless of how careful the experiment is carried out. In addition, the sensitivity of the instruments tends to drop down with use and the signals tend to have strong interferences from the background noises. 


\noindent   Conventional methods for removing noise exist, such as improving accuracy by finding the source of noise and eliminating its effect at the data acquisition stage, or suppressing the noise by replicating the measurements. It can be seen that both of the approaches are practically not feasible. The first approach to find noise in a highly sophisticated instrument has hidden requirements of a significant amount of expertise in that field, while the second is not feasible for financial considerations if samples are of biological, clinical or pharmaceutical origins~\cite{b4}. Therefore, signal processing methods are always needed in such scenarios to denoise pulse-shaped signals.

\noindent  In this paper, we propose to  extend the SCSA method to a more general signal denoising framework and analyze the performance of this approach. To  deal with more general pulse shaped signals, we propose to combine the SCSA with an optimization that computes the optimal values for the semi-classical parameter under some curvature constraints. However and unlike the previous contribution \cite{b1}, which uses prior knowledge on the position and the peaks of the MRS spectrum  and also   the predominance of the noise in some specific areas allowing the computation of the SNR, in this paper we use a curvature constraint that makes the approach applicable for any pulse shaped signal. We refer to the algorithm introduced  in  \cite{b1}  ($\alpha$-SCSA) and the new  proposed algorithm C-SCSA for Curvature-SCSA. 

\noindent  The paper is structured as follows. Section II provides a brief introduction of the SCSA method, and a description of the notation required in the rest of the paper. In Section III the major concepts of curvature denoising as well as conventional $\alpha$-SCSA denoising are described. Consequently, a novel SCSA-based denoising strategy is presented along with other popular denoising techniques. The performance evaluation of the novel denoising technique C-SCSA is  illustrated in Section IV.  The performance of the different SCSA techniques and existing popular denoising methods are illustrated in Section IV, and the final conclusions are drawn in Section V.


\section{SCSA: a brief introduction}
\subsection{SCSA for signal reconstruction} 
\begin{figure}[h] 
\vspace{-6mm}
    \centering
  \subfloat[\label{signal}]{%
       \includegraphics[width=0.4\linewidth]{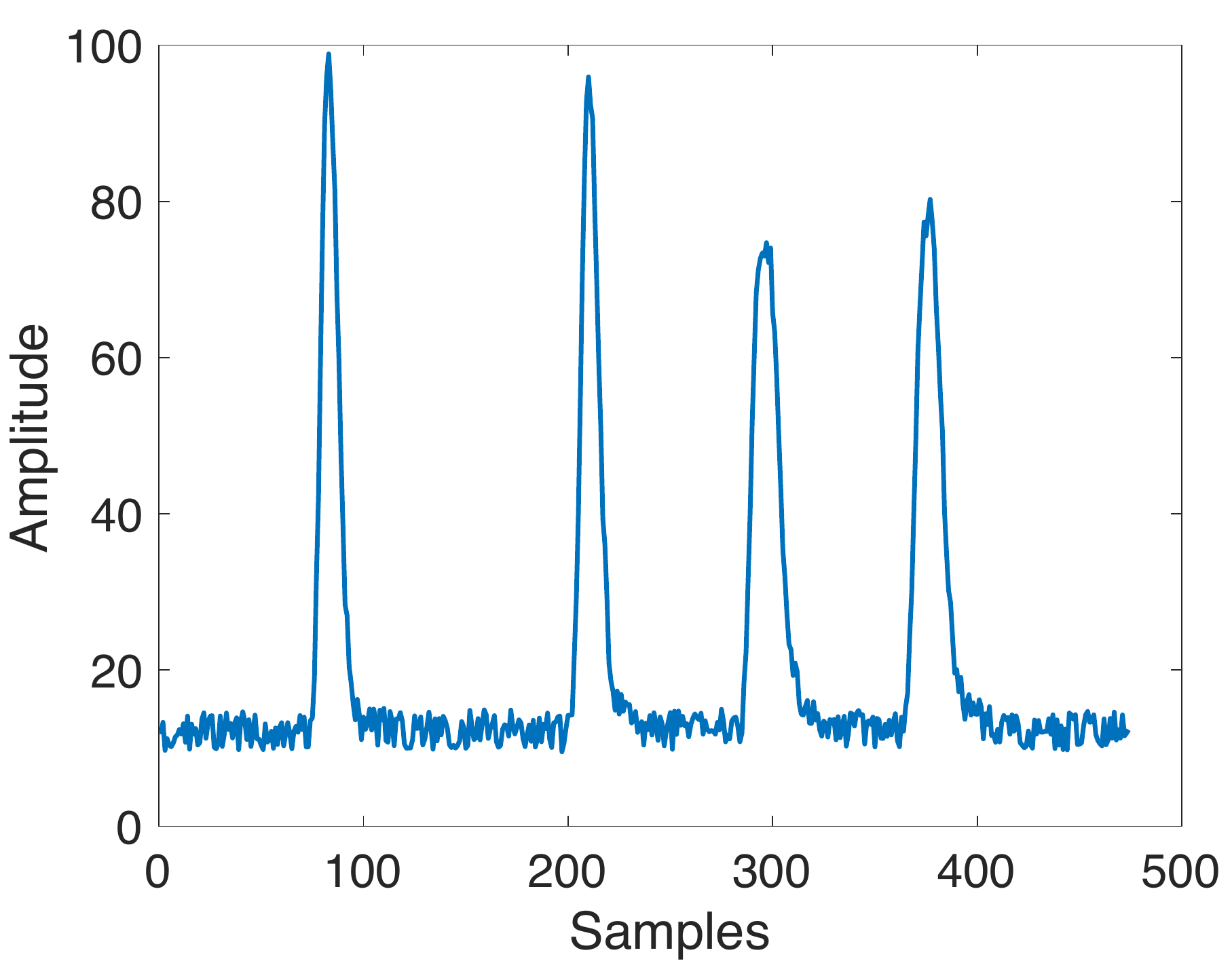}}
    \hfill
  \subfloat[\label{eigenfunction}]{%
        \includegraphics[width=0.4\linewidth]{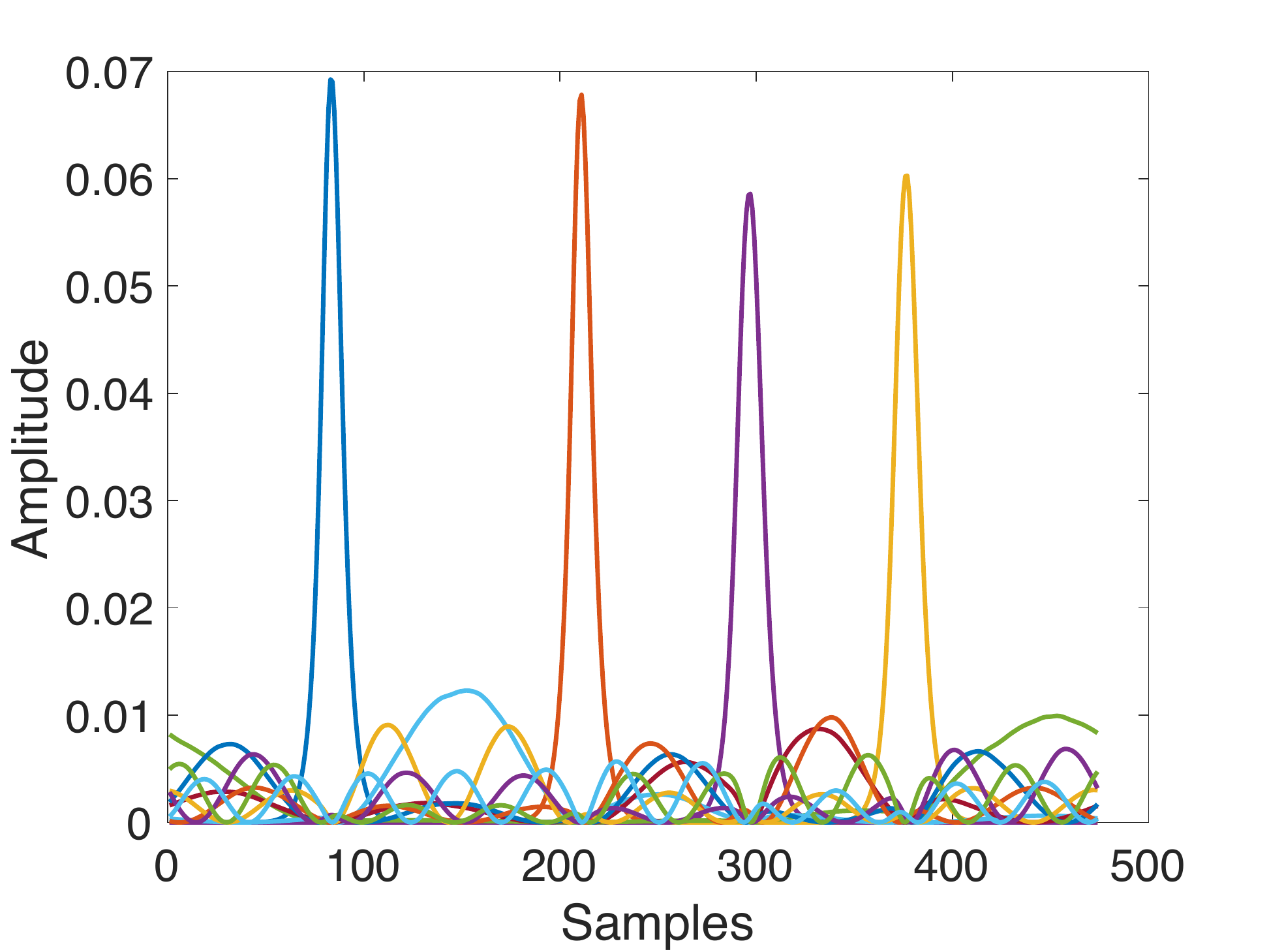}}

  \caption{\textbf{SCSA method}. \textbf{(a)} Input Signal with 4 major peaks \textbf{(b)} Squared eigenfunctions of the signal, with only few of them (4 in this case) corresponding to the major peaks and rest of them explaining the details}
  \label{fig:es} 
\end{figure}

\noindent  The SCSA method decomposes a real positive signal $y(t)$ into a set of squared eigenfunctions through the discrete spectrum of the Schr\"{o}dinger operator. The reconstructed signal $y_h(t)$ is presented by the squared eigenfunctions (refer to ~\cite{b2})
\begin{equation}
y_h(t)=4h\sum_{n=1}^{N_{h}}\kappa_{nh}\psi_{nh}^2(t), t\in\mathbb{R}, \label{eq}
\end{equation}
where $\lambda_{nh}=-\kappa_{nh}^2$, with $\kappa_{1h}> \kappa_{2h}> \cdots>\kappa_{nh}$ are the negative eigenvalues, and \{$\psi_{1h}, \psi_{2h}, \cdots, \psi_{nh}$\} are the corresponding $L_2^2$-normalized eigenfunctions ($n=1, 2, \cdots, N_h$) such that 
\begin{equation}
-h^2\frac{d^2\psi(t)}{dt^2}-y(t)\psi(t)=\lambda\psi(t)\label{equ:sch_equ}.
\end{equation}
\noindent The $L_2^2$-normalized eigenfunctions reconstructed from a pulse shaped signal  are shown in Fig.~\ref{fig:es}. $N_h$ is the number of negative eigenfunctions and $h$ is a positive parameter known as the semi-classical constant. It is found that when $h$ tends to zero, the reconstructed spectrum $y_h$ converges to the true spectrum $y$. This matches the  semi-classical properties of the Schr\"{o}dinger operator where the number of negative eigenvalues thus the number of corresponding eigenfunctions increases when $h$ decreases. One of the important characteristics is that eigenfunctions which correspond to large eigenvalues  represent  the profiles of the peaks, whereas the remaining functions characterize the noise details of these profiles. Fig.~\ref{fig:es}(b) shows an example where \{$\psi_{1h}, \cdots, \psi_{4h}$\} correspond to signal's  major peaks. The SCSA analyzing process used in this paper is the standard one. According to this procedure, the SCSA algorithm decomposes the signal as follows.

\begin{itemize}
\item[1] Compute the eigenvalues and the eigenfunctions of the discrete
Schr\"{o}dinger operator using Equation (2) for a given value of $h$ by
solving the eigenvalue problem of the matrix, $-h^2 D-Y$, where $D$
represents the differentiation matrix of the Laplacian, computed
using a Fourier pseudo‐spectral method, and $Y$ is a diagonal
matrix whose entries are the values of the noisy signal $y(t)$.
\item[2] Find the negative eigenvalues and the associated eigenfunctions
for the matrix above.
\item[3] Normalize the eigenfunctions and compute $y_h$ using Equation (1).
\item[4] If the reconstruction is accurate, stop; if not, decrease the value of
$h$ and go to step 1.
\end{itemize}

\subsection{SCSA for signal denoising} 
\noindent Now, let's consider the following noisy signal 
\begin{equation}
y_\delta(t)=y(t)+n(t)\label{equ:sch_equ},
\end{equation}
where $y(t)$ is the noiseless signal and $n(t)$ is  the additive noise. The aim of digital signal denoising is to produce an accurate estimate  of   the original signal $y(t)$. The  noise variance of $n(t)$, which can be known or unknown depending on different cases, is denoted as $\sigma$

 \noindent As in the reconstruction, the parameter $h$ plays a key role in the denoising with the  SCSA method. On one hand, when $h$ tends to zero, the reconstructed spectrum $y_h$ converges to the noisy signal  $y_\delta$. On the other hand, It is demonstrated that $N_h$ increases when $h$ decreases and the squared eigenfunctions $\psi_{nh}$ are such that  the number of oscillations in the eigenfunctions increases with the order $n$ while their amplitude decreases. Therefore,  the highest order eigenfunctions will  mainly reconstruct the noise. 
 
\begin{figure*} [htbp]
\vspace{-6mm}
    \centering
  \subfloat[$h=100$, $N_h=6$\label{1a}]{%
       \includegraphics[width=0.333\linewidth]{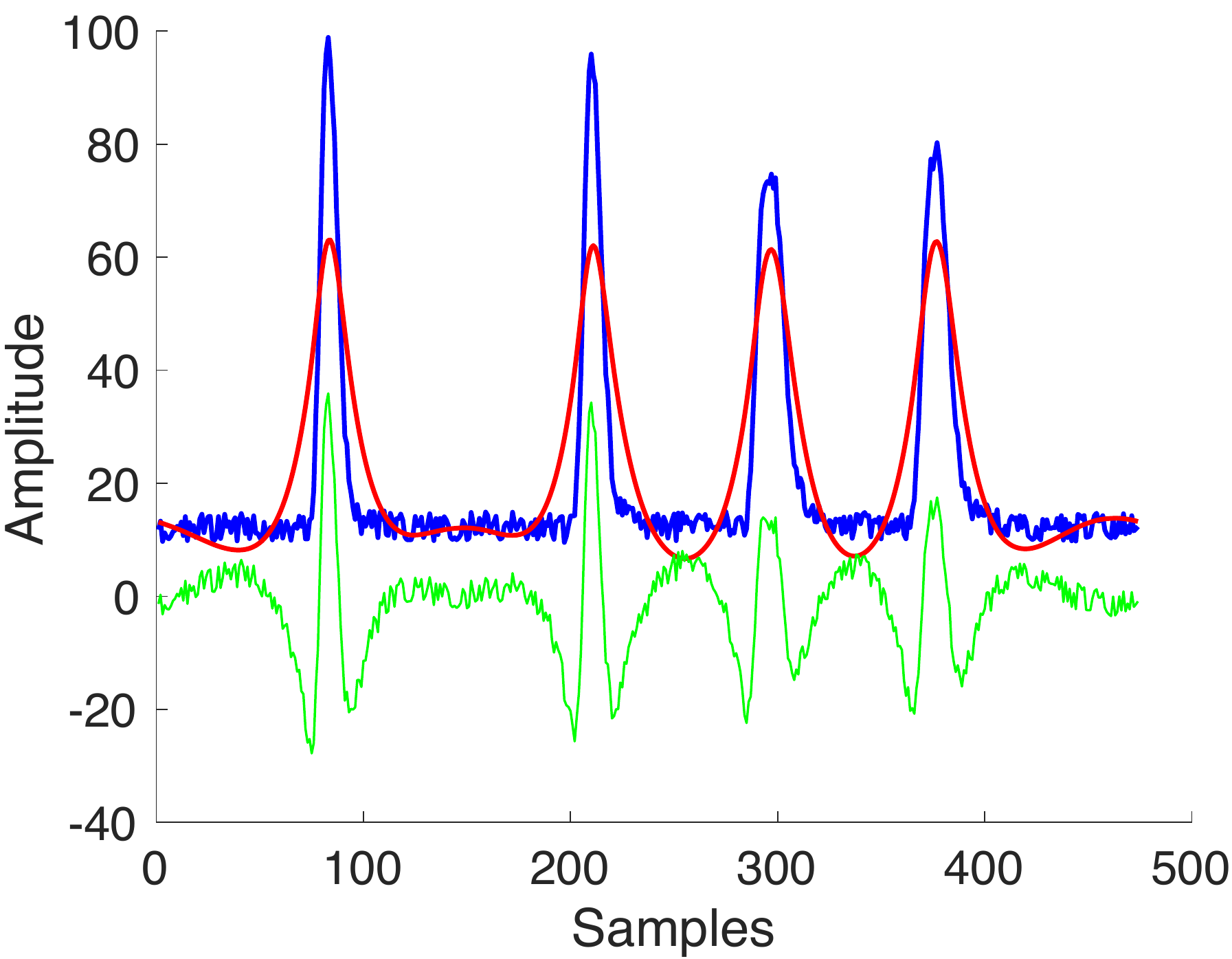}}
    \hfill
  \subfloat[$h=50$, $N_h=13$\label{1b}]{%
        \includegraphics[width=0.333\linewidth]{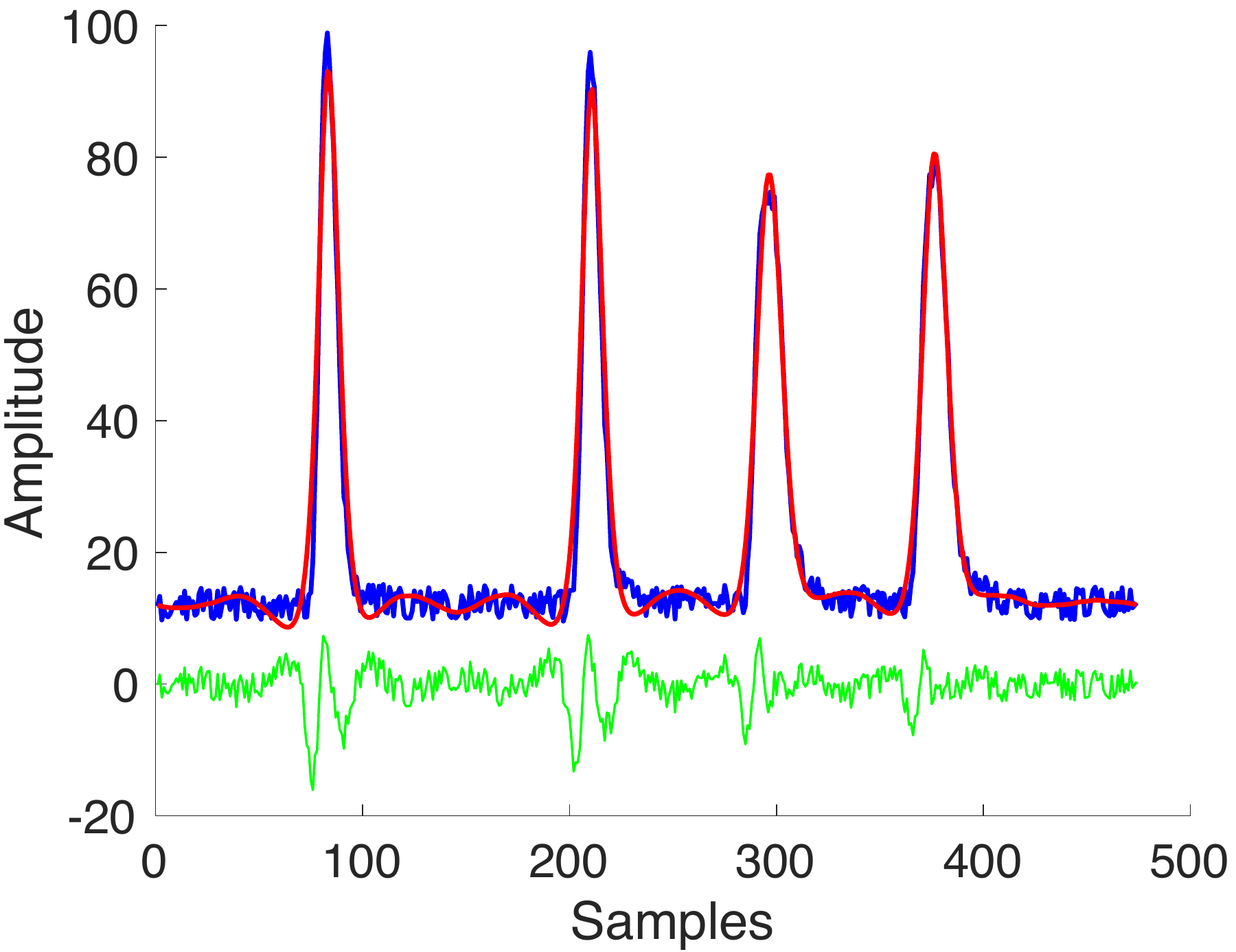}}
            \hfill
  \subfloat[$h=35$, $N_h=18$\label{1c}]{%
        \includegraphics[width=0.333\linewidth]{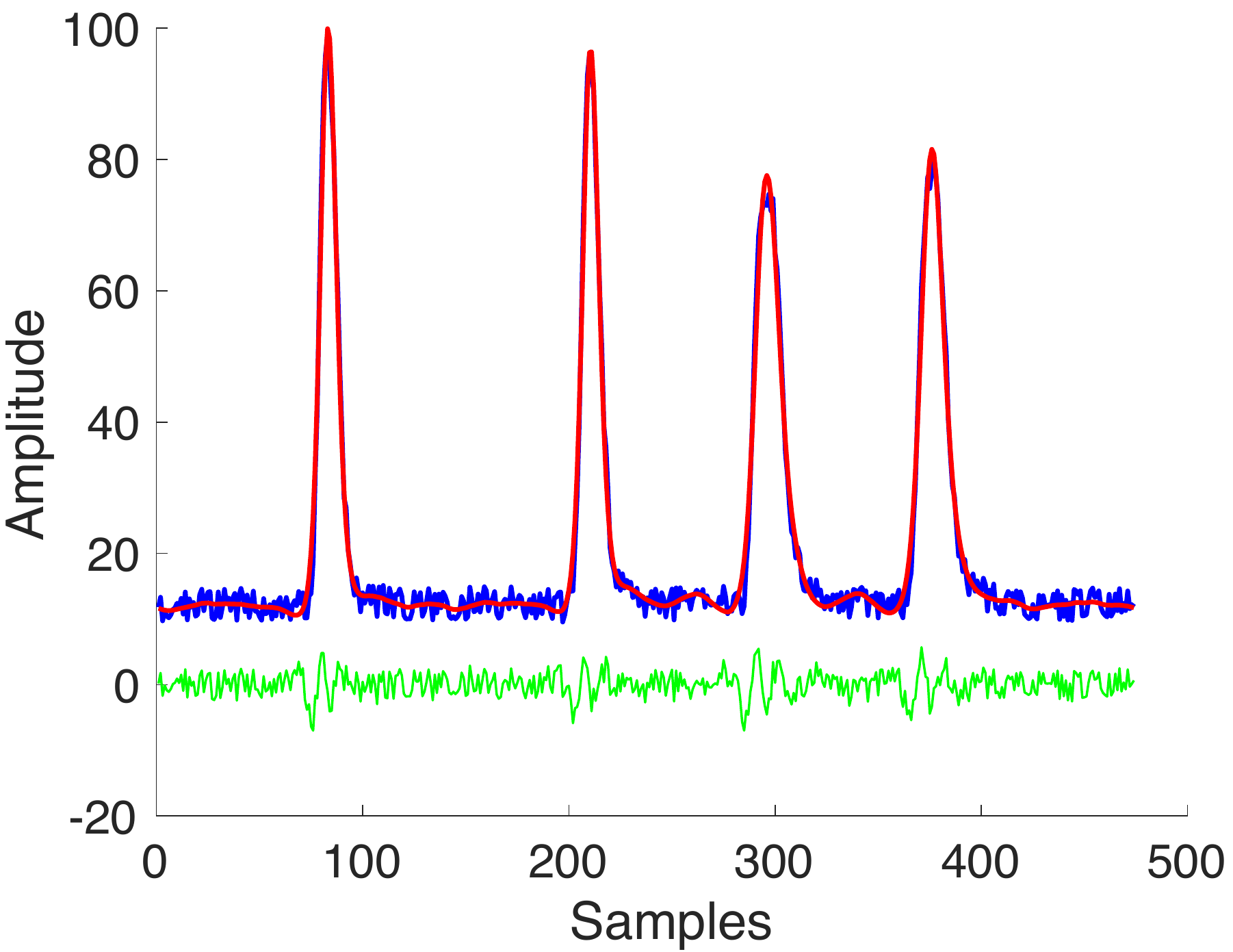}}
            \hfill
  \subfloat[$h=15$, $N_h=43$\label{1d}]{%
        \includegraphics[width=0.333\linewidth]{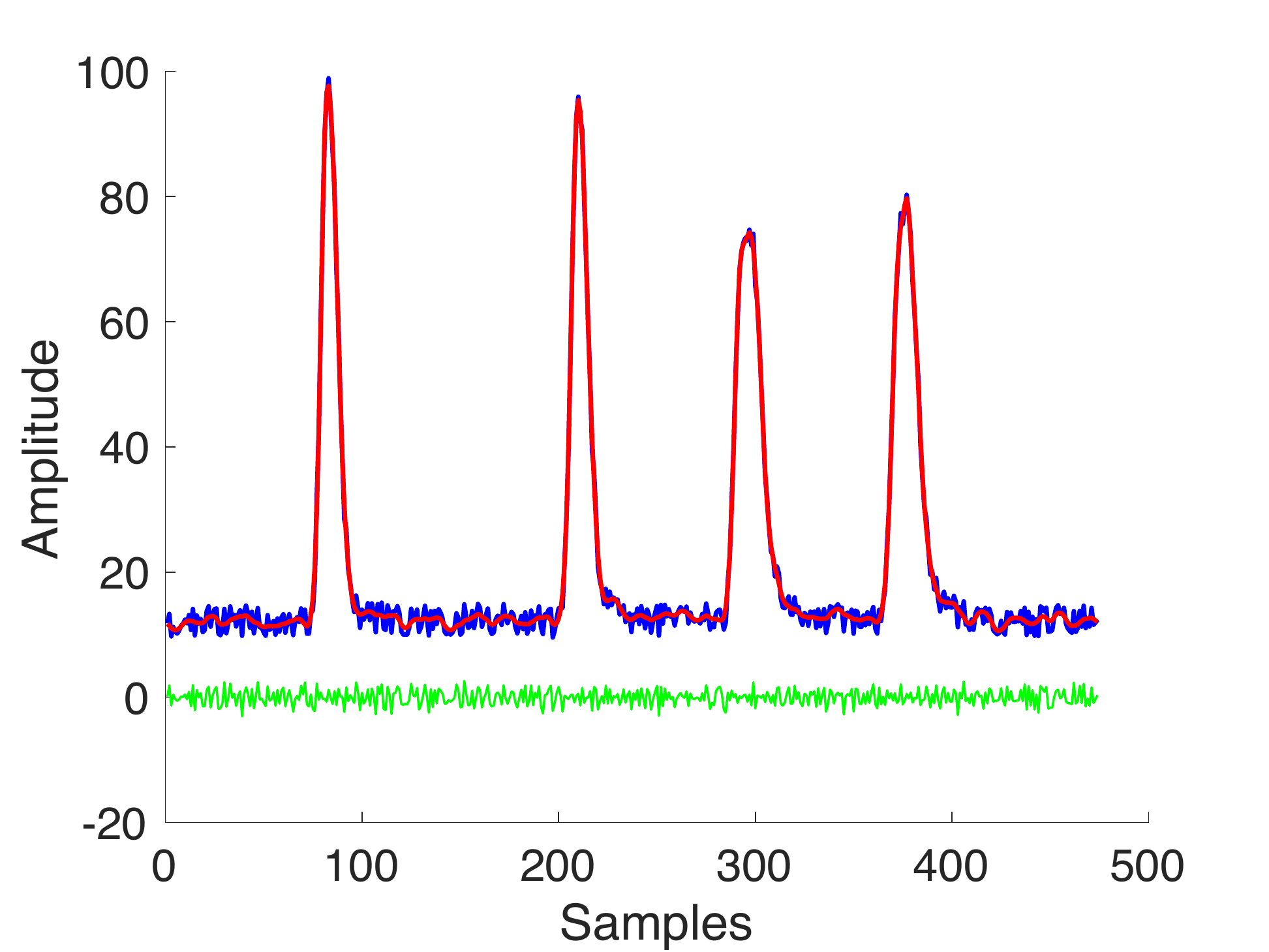}}
                    \hfill
  \subfloat[$h=5$, $N_h=127$\label{1e}]{%
        \includegraphics[width=0.333\linewidth]{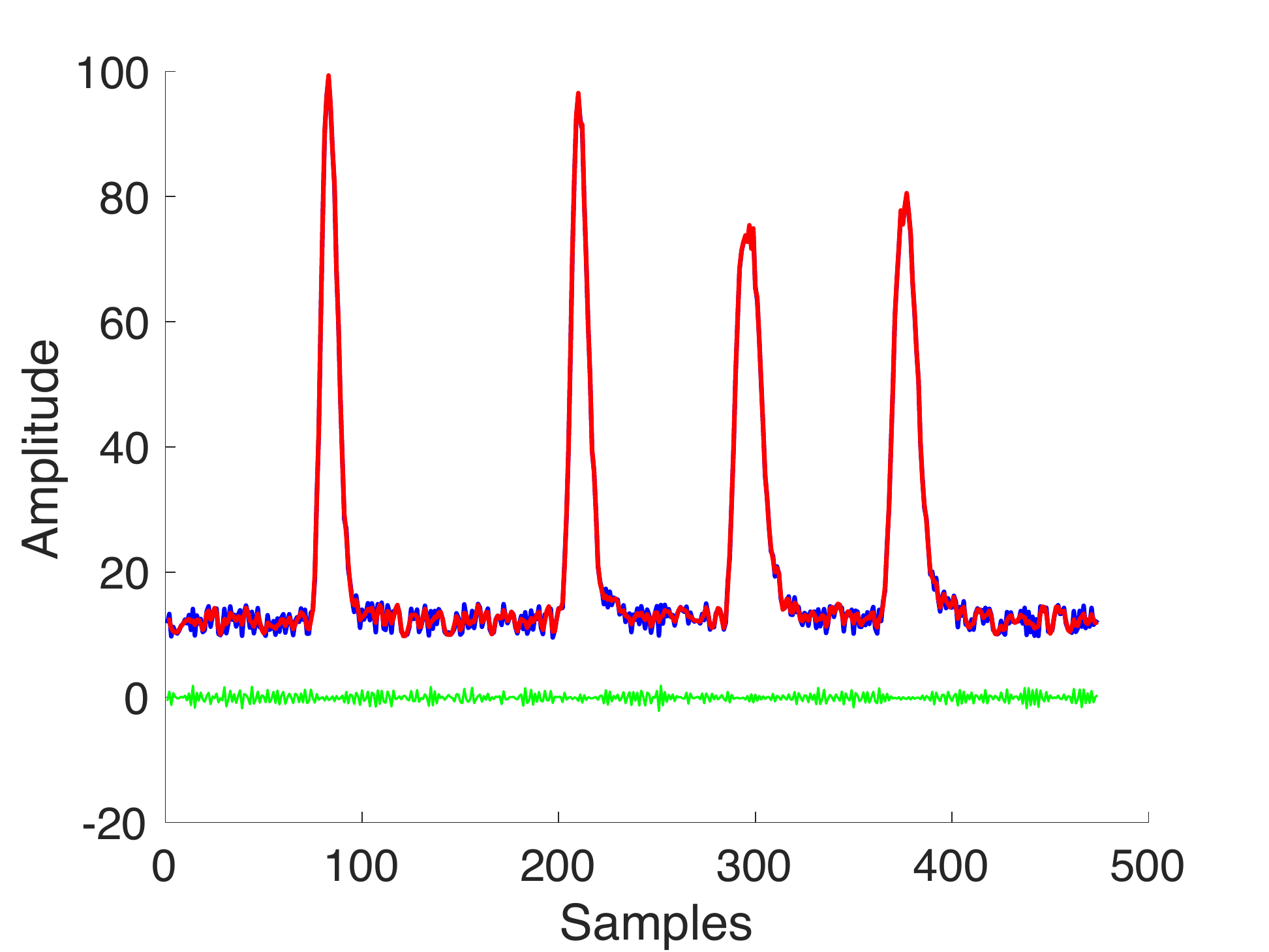}}
         \hfill
  \subfloat[$h=3$, $N_h=212$\label{1f}]{%
        \includegraphics[width=0.333\linewidth]{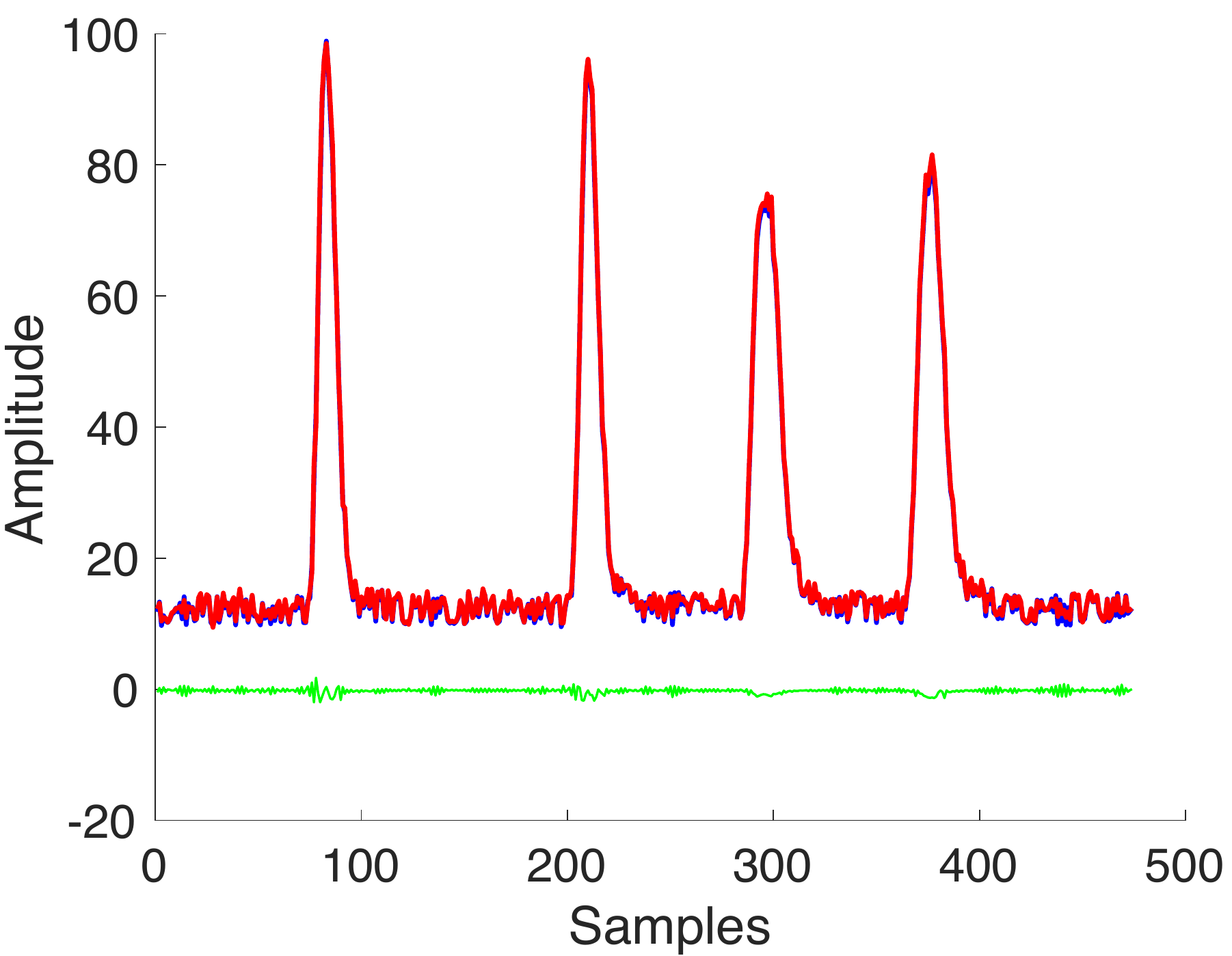}}
  \caption{\textbf{SCSA's application in signal denoising.} Input signal with noise(in blue), SCSA spectrum (in red) and residual (in green) \textbf{(a)} Small $N_h$ value, not capable of reconstructing the signal. \textbf{(b)(c)} $N_h$ increases, the peaks of the signals are recovered, without recovering the rest part of the signal. \textbf{(d)} With $N_h$ continuing to increase, the non-peak areas are also recovered, while the noise is separated. The corresponding $h$ is a suitable denoising coefficient.  \textbf{(e)(f)} High $N_h$ value will recover the whole signal, including the original signal and noise.}
  \label{fig:cossfun} 
\end{figure*}

  \noindent In a general sense, as $N_h$ value increases, both the original signal and noise will be gradually reconstructed, first the major peaks and details of the signal and then the noise, as shown in Fig.~\ref{fig:cossfun}. A selection of optimal $h$ has to be made in order to separate noise from the original signal. In simulation, $h$ is initiated at a relatively small value at first and then gradually increased to discard the noise part. One can infer that the choice of the stop criterion is critical, since it sets the optimum $h$ value, which leads to a reliable signal reconstruction and therefore to an accurate data analysis. It is found that the best stop criterion is a function not only of the minimum reachable distance between $y_\delta(t)$ and reconstructed signal $y_h$, but also of the Signal to Noise Ratio (SNR) of $y_h$ reconstructed spectrum \cite{b1}:
 \begin{equation}
J=\sum_{i=1}^M||y_{\delta}-y_h||_{peaki}^2+\frac{\alpha}{|\textrm{SNR}_{y_h}|}, 
\end{equation}
where $M$ represents the number of
peaks and $\alpha$ is a positive weight parameter, which  allows to balance  accuracy and denoising.  $ \textrm{SNR}_{y_h}$
is the SNR of $y_h$ computed using the following formula 
$$
\text{SNR}_{y_h} = \frac{\max\{ |y_h | \}}{\mbox{std} \{ y_h |_{[t_1,t_2]} \}}.
$$ 
where the interval $[t_1,t_2]$ is the interval where the noise is dominant, $\max$ and  $\mbox{std}$ represent the maximum and standard deviation of the function $y_h$

\noindent This type of cost function is quite standard, and keeps a balance
between fidelity to the signal and denoising effect. However, this method has some pre-assumptions, in which signal peak localization is needed and also the interval where the noise is dominant needs to be known.

\section{Curvature based SCSA Denoising}
\noindent In this section we propose a new denoising algorithm based on the SCSA and some curvature constraints. We refer to this algorithm C-SCSA. The C-SCSA proposes  a general solution to reduce signal noise  without much pre-knowledge of the characteristics of it. While the general principles of denoising is still the same, i.e. to come up with a standard of properly select $h$ in order to reduce the noise of  the signal, the cost function selection will need more mathematical intuition. 

\noindent Therefore,  inspired by the smoothing methods of~\cite{b8}, we propose a cost function $J$ in the following form: 
\begin{equation}
J=||y_{\delta}-y_h||_2^2+\mu \int \! |k(t)| \, \rm d t, 
\end{equation}
 where $|k(t)|$ is a certain smoothness penalty term which operates on the reconstructed signal $y_h$. $\int \! |k(t)| \, \rm d t$ describes the "wiggliness'' of the reconstructed signal $y_h$,  $\mu$ is a non-negative smoothing parameter that needs to be properly selected. It depends on the characteristic of the input signal. Larger values of $\mu$ force $y_h$ to be smoother. In this paper we define smoothness penalty term $k(t)$ to be the curvature.
\begin{equation}
k(t)=\frac{|y_h^{\prime\prime}(t)|}{(1+y_h^{\prime}(t)^2)^{\frac{3}{2}}}\label{eq:k1}.
\end{equation}
Let $(y_1, y_2, \cdots, y_N)$ to be input signal $\textbf{y}_{\delta}$ with $N$ samples, with each sample taken as separate random variables. Let $x_m=y_{m+1}-y_m$ and $w_m=y_m-y_{m-1}$, $m=2, 3, \cdots, N-1$. Without loss of generality, let's assume $x_m$ and $w_m$ to be jointly Gaussian and zero mean with variance $\sigma_m^2$, with their joint distribution defined as:
\begin{equation}\label{eq:e10}
f(x_m,w_m) = \frac{\exp(-\frac{x_m^2+w_m^2-2\rho_m x_mw_m}{2\sigma_m^2(1-\rho_m^2)})}{2\pi\sigma_m^2\sqrt{1-\rho_m^2}}, 
\end{equation}
where $\rho_m = \frac{COV\{x_m,w_m\}}{\sigma_m^2}$. If the signal are equally spaced with interval denoted as $\Delta$, the curvature defined in Eq.~\eqref{eq:k1} can be approximated as Eq.~\eqref{eq:e11} at the $m$th sample :

\begin{equation}\label{eq:e11}
k_m = \frac{|x_m-w_m|}{\Delta^2(1+\frac{(x_m+w_m)^2}{4\Delta^2})^{\frac{3}{2}}}. 
\end{equation}
\newtheorem{theorem}{Proposition}
\begin{theorem}[Expectation of the curvature]\label{theorem}
Given $f(x_m,w_m) $ and $k_m$ in Eq.~\eqref{eq:e10} and \eqref{eq:e11}, $E\{k_m\}$ can be approximated as:
\small{\begin{align} \label{eq:e1}
E\{k_m\} 
&= \int_{-\infty}^{\infty}\int_{-\infty}^{\infty} \!  k_m f(x_m,w_m) \, \dd x_m \dd w_m \nonumber\\
&=  \frac{4}{\pi \Delta}\sqrt{\frac{1-\rho_m}{1+\rho_m}}
\int_{0}^{\infty} \! \frac{1}{(1+\eta^2)^{\frac{3}{2}}}
\exp\left(-\frac{\Delta^2}{\sigma_m^2(1+\rho_m)}\eta^2\right) \, \dd \eta
\end{align}}
Note that $k_m>0$, with $|\rho_m|\ll1$ in most cases. 
\end{theorem}
The proof of the theorem is given in the appendix.
 Since $x_m$ and $w_m$ are subtraction between two neighborhood samples, they resemble noise distributions in homogeneous regions of the signal (a common assumption for noise propagation analysis like~\cite{b9}), whose distribution $\sigma_m^2$ are of the similar distribution of the noise, i.e. $\sigma_m^2\simeq2\sigma_{noise}^2$ when noise is identically distributed.
Therefore when noise level $\sigma_{noise}$  increases, the $\sigma_m^2$ also increases. Given Eq.~\eqref{eq:e1} one can easily infer that when $\sigma_m^2$ increases, $E\{k_m\}$ also increases. Therefore, in order to reduce noise level, we use curvature term $k$ in cost function $J$. Then, given $N$ samples of the signal  $y_{\delta}$, we propose a scanning method which iteratively scans $h$, to minimize the cost function $\bar{J}$  
\begin{equation}
\bar{J}=\sum_{i=i}^{N}[y_{\delta}(t_i)-y_h(t_i)]^2+\mu \sum_{i=1}^N k(t_i)\label{equ:J}
\end{equation}
where  $y_{\delta}$ is the input noisy signal and $y_h$ is the signal reconstructed with SCSA method. $N$ is the number of samples in $y_\delta$. $\mu$ is automatically adjusted given a specific type of signal following $\frac{\max\{|y_\delta(t_i)|\}}{\sum k(i)}\cdot10^\nu$ ($\nu \in \mathbb{Z}$, $i=1,2,\cdots,N$) in order to make sure that fidelity term and penalty term are of similar order of magnitude.  The algorithm makes a step further comparing to~\cite{b1} in that it doesn't need to locate the signal peaks during the denoising process, which in some cases is not  trivial. The fidelity term in the cost function $\bar{J}$ ranges the whole input noisy signal. Also, the SNR is unknown in many cases, while curvature is a characteristic of all types of signals that can be numerically computed. 

\section{Simulation Results}
\subsection{Denoising methods  }
\noindent There are many existing methods for pulse-shaped signal denoising. With respect to traditional digital filters, moving average filter is the simplest method~\cite{b10,b11}, with the drawbacks of temporal autocorrelation at a lag determined by the length of the moving window, thus shifting the peak position which will negatively affect the signal. In comparison, the Savitzky-Golay algorithm (SG)  is more effective in smoothing noisy data obtained from spectrum data (for example~\cite{b12}) and is currently the  most commonly applied filter to eliminate the irrelevant information from noisy input data~\cite{b13}. Other well-known candidates for denoising  are techniques based on wavelets\cite{b16}. There are also other empirical methods such as Empirical Modes Decomposition (EMD) method  \cite{b18} that we include in our comparison in this paper.  The reason is that both EMD and SCSA share some common properties. Indeed, both of them adapts  flexible basis during denoising: EMD approach  thresholds the Intrinsic Mode Functions (IMFs) and the SCSA method  thresholds the number of squared eigenfunctions.

\subsection{Simulated Signal and Synthetic Noise}
\noindent  Spectral peaks can be modeled by Gaussian peaks, Lorenz peaks or their combination~\cite{b14}. We choose multi-peak Gaussian signals as test data. The Gaussian peaks are generated by
\begin{equation}
G_s(t)=\displaystyle\sum_{i=1}^{\bar{M}}A_i \exp\left[-\frac{(t-u_i)}{2\sigma_i^2}\right],
\end{equation}
$\bar{M}$ is the number of peaks, $A_i$, $u_i$ and $\sigma_i$ is the amplitude, position and width of peak $i$ ($i=1,2,3,\cdots,\bar{M}$). The noise added to the signal is Gaussian noise generated by \textit{rand()} function in Matlab. In the simulation, $\bar{M}=1, u_1=5, \sigma_1=15, A_1=2$. Noise levels vary between 1\% and 12\% with 0.5\% interval.

\paragraph{Peak preserving performance:} 
\noindent  As an evaluation of peak preserving performance of different methods, we use a  simulated Gaussian peak to assess peak-preserving performance. $\mu$ is selected according to SCSA method. Depending on the level of noise, $\mu$ is usually selected to be smaller when SNR is smaller. In the Savitzky–Golay method, the width of the sliding window is 29 points and polynomial degree is 4. In wavelet method, the function “wden” in Matlab toolbox was used to smooth the signal, the wavelet is \textit{Sym4} and decomposed level is 3. We compare the performance of peak-preserving which is described with peak height.  The peak height relative error is determined  as follows:
\begin{equation}
\textit{Error}_M=\frac{abs(M_h-M_c)}{M_c}\times 100,%
\end{equation}
where $M_h$ is   the peak amplitude of denoised signal and  $M_c$ peak amplitude of clean signal.
\noindent Denoting  the peak width of denoised signal $W_h$ and peak width of clean signal $W_c$, the formula of peak width relative error is given by:
\begin{equation}
\textit{Error}_W=\frac{abs(W_h-W_c)}{W_c}\times 100%
\end{equation}

\begin{figure} [htbp]
\vspace{-6mm}
    \centering
  \subfloat[\label{signal}]{%
       \includegraphics[width=0.5\linewidth]{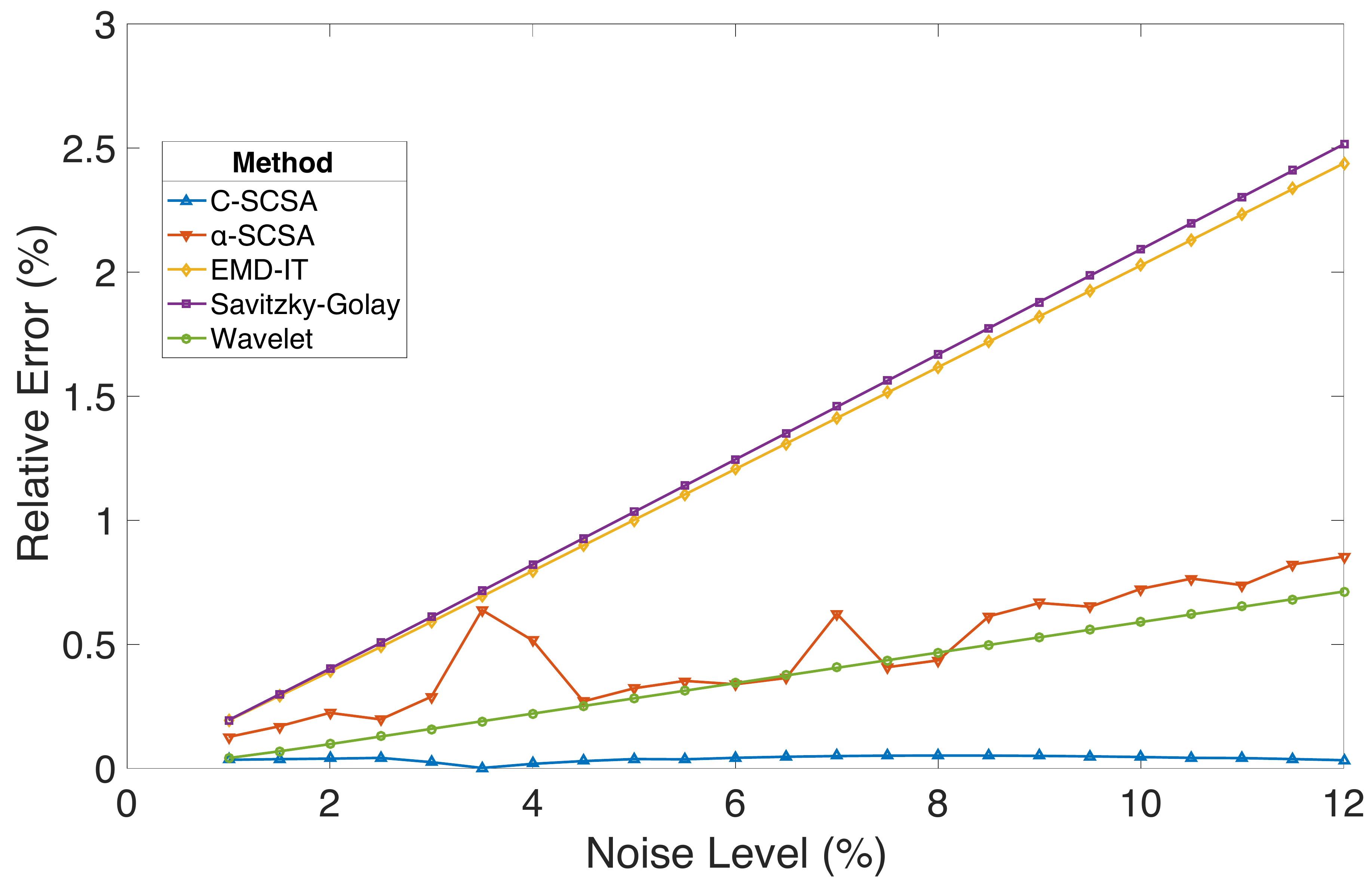}}
    \hfill
  \subfloat[\label{eigenfunction}]{%
        \includegraphics[width=0.5\linewidth]{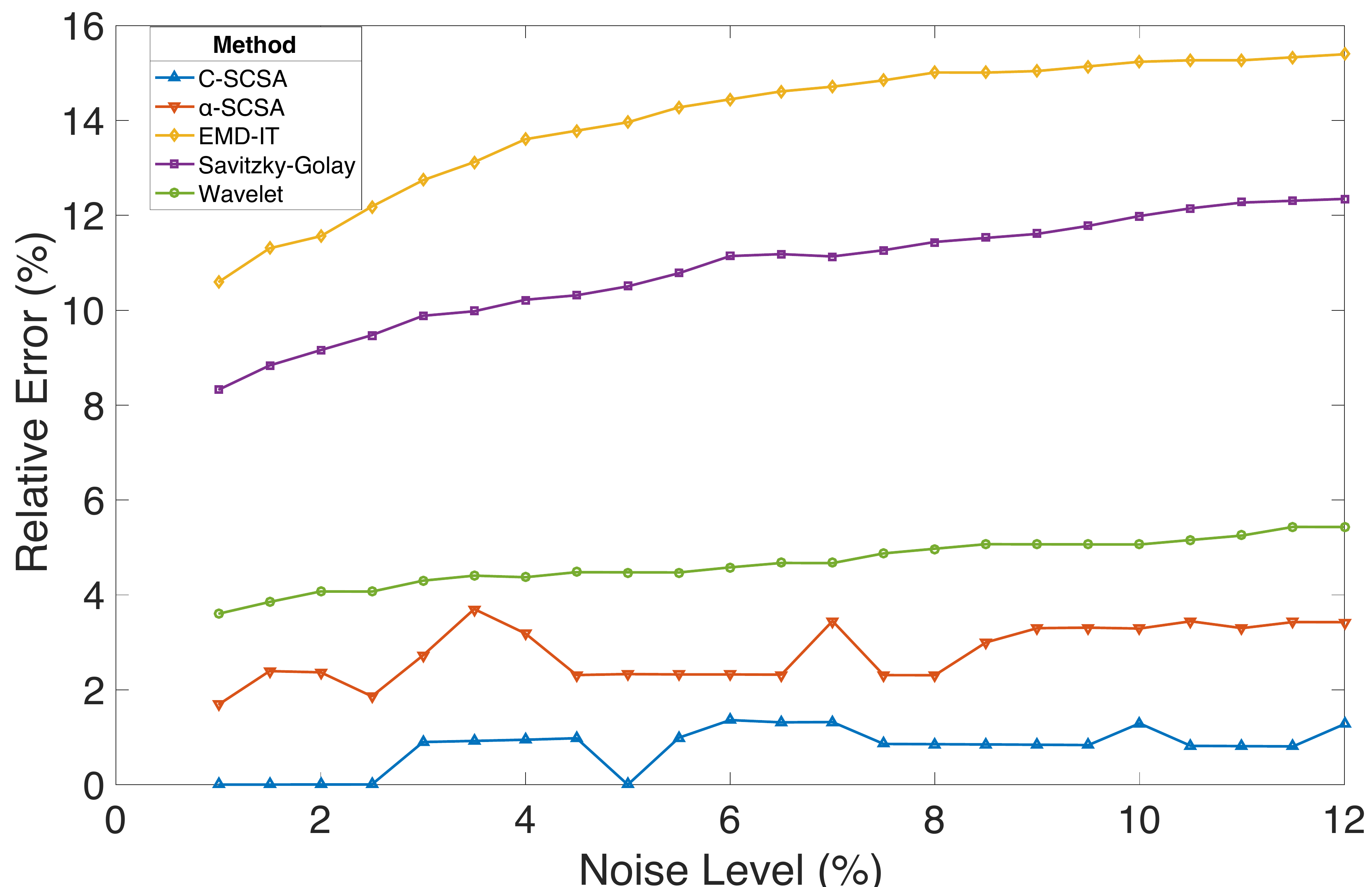}}

  \caption{\textbf{Peak preserving performance for single Gaussian peak }. \textbf{(a)} Peak height preserving performance  \textbf{(b)} Peak width preserving performance (Noise level ranges between 1\% and 12\% (with interval 0.5\%) as shown in the horizontal axis).}
  \label{fig:e1s} 
  \vspace{-2mm}
\end{figure}

\noindent We compare the performance of peak-preserving. in different noise conditions. The results are shown in Fig.~\ref{fig:e1s}.  One can easily see that the relative errors of peak heights increase as noise level increases. At the same time, one can also see that the SCSA method has better performance of peak-preserving than other methods in terms of peak preserving performance.
 
 \paragraph{Denoising Performance and comparison: } 
 The denoising performance analysis throughout this paper is assessed by the mean squared error (MSE):
 $$
\text{MSE} = \frac{1}{N}\displaystyle\sum_{i=1}^N (y(t_i)-y_h(t_i))^2,
$$
and the signal to noise ratio (SNR):
$$
\text{SNR} = 10\log\frac{\displaystyle\sum_i^N y_h(t_i)^2}{\displaystyle\sum_i^N(y(t_i)-y_h(t_i))^2},
$$
where $y$ is the original signal with length $N$, $y_h$ is the related denoised signal.

\noindent  In this experiment, a 5-peak simulated signal is generated to demonstrate the denoising process. To make the signal more general, peaks are with different widths and heights with certain peaks overlapping each other.

\small{
\begin{landscape}

\begin{table*} [htbp]
\caption{Real noise experiment carried out over several records from the MIT BIH arrhythmia database.}\label{table}
\begin{tabularx}{1.5\textwidth}{@{}l*{14}{C}c@{}}
\toprule
     &   &SNR=7dB  &&&  && SNR=10dB&&&&&SNR=14dB& &&\\ 
     \cline{2-5} \cline{7-10} \cline{12-15}
     & \small{SNR\textsubscript{SCSA}$^{\mathrm{a}}$} & \small{SNR\textsubscript{SG}$^{\mathrm{b}}$ }& \small{SNR\textsubscript{wt}$^{\mathrm{c}}$}&  \small{SNR\textsubscript{EMD}$^{\mathrm{d}}$} && \small{SNR\textsubscript{SCSA} }& \small{SNR\textsubscript{SG} }& \small{SNR\textsubscript{wt}} &  \small{SNR\textsubscript{EMD}} && \small{SNR\textsubscript{SCSA} }& \small{SNR\textsubscript{SG}} & \small{SNR\textsubscript{wt} }&\small{SNR\textsubscript{EMD}}\\ 
\hline
100  & 7.9186   & 6.7532        &8.5615  &7.1337   &  & 11.0174    & 8.2943 &11.5837&10.2958 && 14.4794    &14.0068      &15.4265   &14.3136& \\ 
103  & 7.7257    &  7.0795     & 8.2101  &   7.0291   & &  10.3102   & 10.0839        &11.1383&10.0694& & 14.6304   &  14.0578        &15.1521 &14.1622&\\

201& 8.6908&7.1139&8.3935&7.4150&& 11.3994&10.1341&  11.3172
& 10.5668&&14.6393&     14.0654
&14.9216& 14.0616\\
202&9.3567&7.1112&8.4234&8.2073&&11.5873&10.0866&11.3181& 10.5834&& 16.5448&14.9503&16.1262   &16.4202\\
222&8.0961&7.0519&8.2660&7.0208&&10.7597&10.0100&11.0539&10.0497&& 14.7465& 14.7333&15.4538   & 15.0091\\
\bottomrule
\end{tabularx}

\vspace{2mm}

$^{\mathrm{a}}$SNR value for the SCSA method,  $^{\mathrm{b}}$SNR value for the Savitzky-Golay filtering method,  $^{\mathrm{c}}$SNR value for the wavelet based method, $^{\mathrm{d}}$SNR value for the EMD-IT based method.
\end{table*}

\end{landscape}
}

\noindent  For Savitzky-Golay filter, signal becomes increasingly smooth as the window size increases. On the other way, too broad of a window will reduce the effect of the resolution enhancement and distort the derivative spectra. The best parameters for the Savitzky-Golay method are selected usually by a trial-and-error method~\cite{b13}~\cite{b15}.  In terms of wavelet method, \cite{b16} presents a selection procedure of mother wavelet basis functions applied for denoising of the noisy signal in wavelet domain while retaining the signal peaks close to their full amplitude. The universal threshold selection by Donoho and Johnstone is applied with varying wavelet basis function, which will also be compared in the following section.

\begin{figure}  [htbp]
\vspace{-6mm}
    \centering
  \subfloat[MSE\label{1a}]{%
       \includegraphics[width=0.5\linewidth]{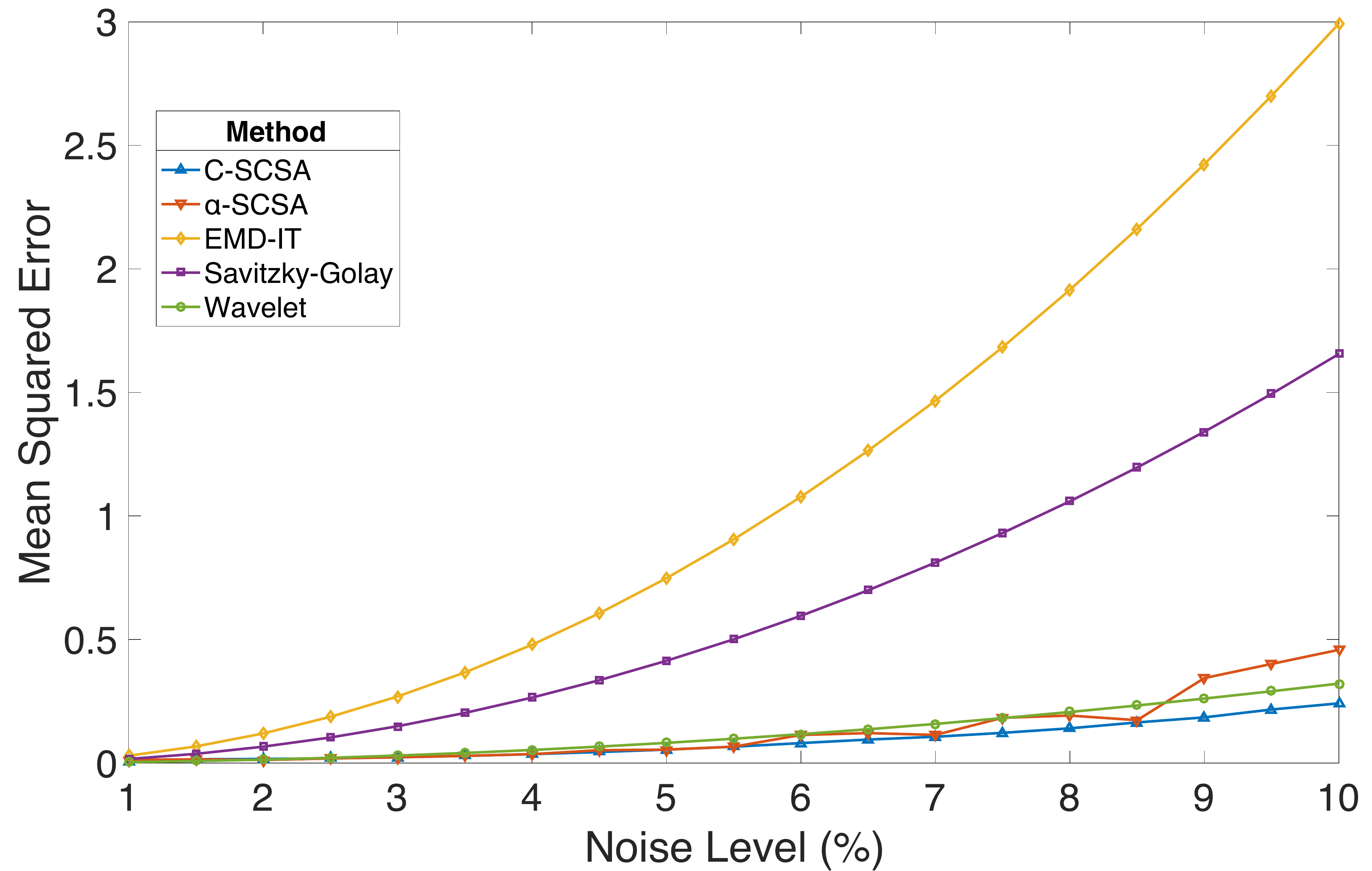}}
    \hfill
  \subfloat[SNR (dB)\label{1b}]{%
        \includegraphics[width=0.5\linewidth]{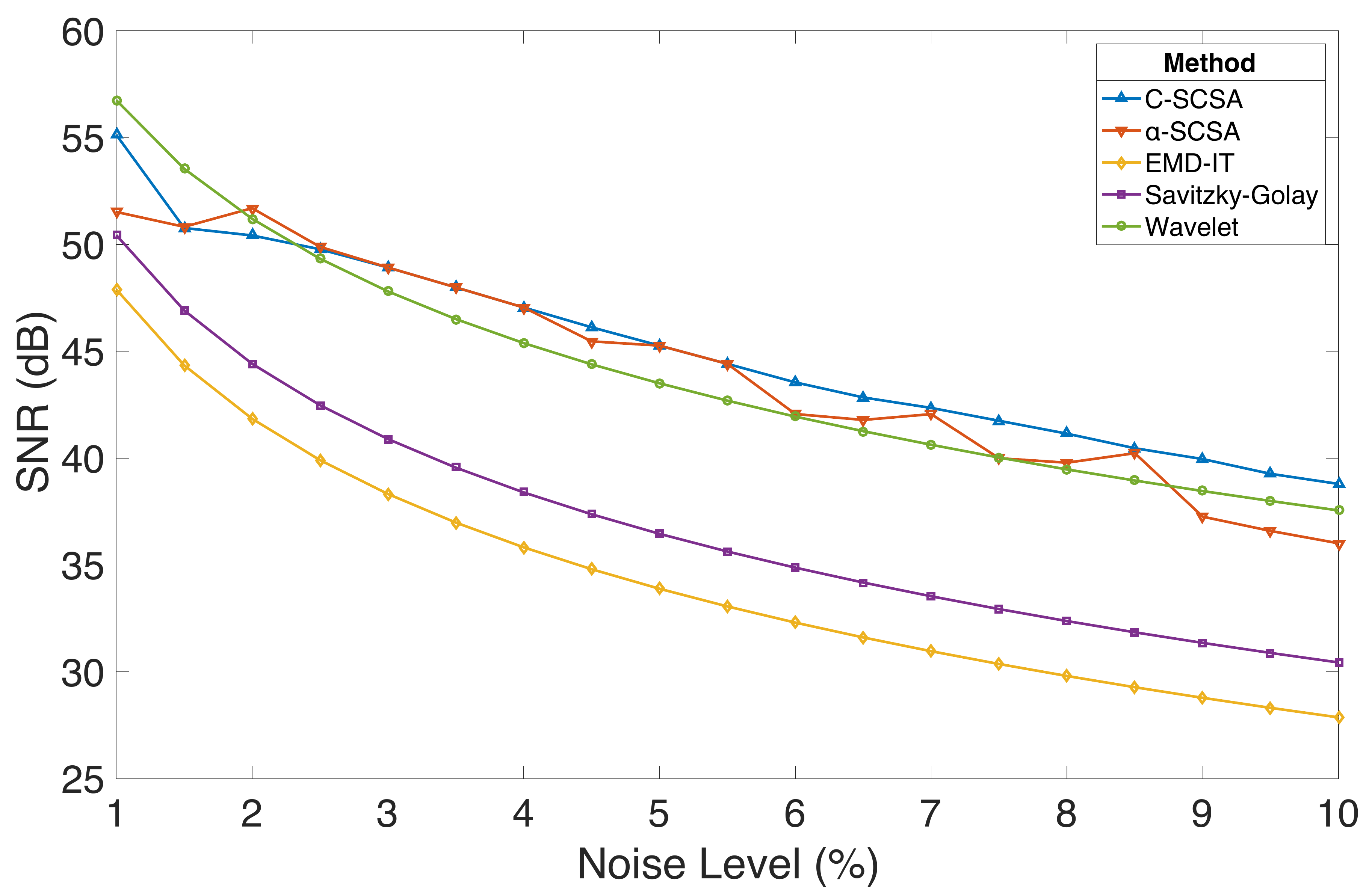}}
\caption{\textbf{Quantitative Denoising Performance of Different Noise Levels.} Noise level ranges between 0.1\% and 15\% (with interval 0.1\%) as shown in the horizontal axis. \textbf{(a)} Mean Squared error . \textbf{(b)} SNR (dB). Both wavelet and Savitzky-Golay methods are optimized as described. \textit{sym4} base function with decomposition level 3 is selected for wavelet method, and order 4 with filter length 17 is selected for Savitzky-Golay method.}
\label{fig:quantitative} 
\vspace{-4mm}
\end{figure}

\noindent  In order to select the best parameter for each method, we optimize each method's parameter at noise level 5\%, where we iteratively optimize its parameter using the noisy signal and true signal. For SG method, consider it having $l$ filter length and $r$ order polynomial,  all possible combinations $(l, r)$ are then tried to yield the best smoothing performance. For wavelet method, method is chosen according to~\cite{b16}, where different base functions are compared and selected. $\mu$ in SCSA cost function is also selected in a scanning manner. We then fix the parameter and try different noise levels. It can be seen from Fig.~\ref{fig:quantitative} that both wavelet method and SCSA based method is out performing the traditional digital filter, while SCSA and wavelet methods are giving comparable  denoising results.

\subsection{ECG Signal Denoising}
\begin{figure}[htbp]
  \centering
\begin{tikzpicture}
\node[anchor=south west,inner sep=0] (image) at (0,0) 
{\begin{overpic}[width=0.8\textwidth]{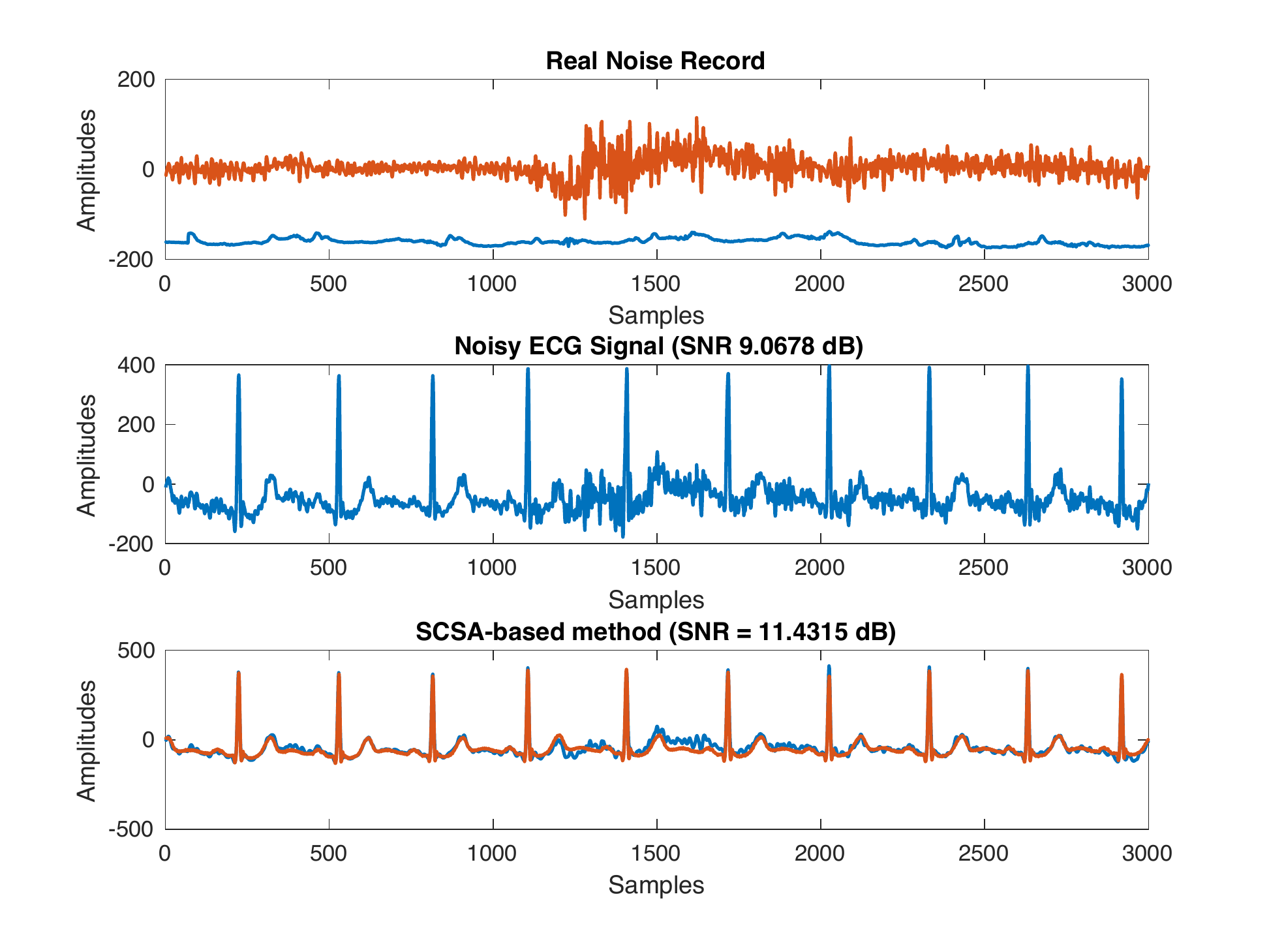}
\end{overpic}};
\begin{scope}[x={(image.south east)},y={(image.north west)}]
\draw[color=black] (0.715,0.19) rectangle ++(0.035,0.12);
\node[anchor=south west,inner sep=0] (sub) at (0.825,0.03) 
{\begin{overpic}[width=0.1\textwidth]{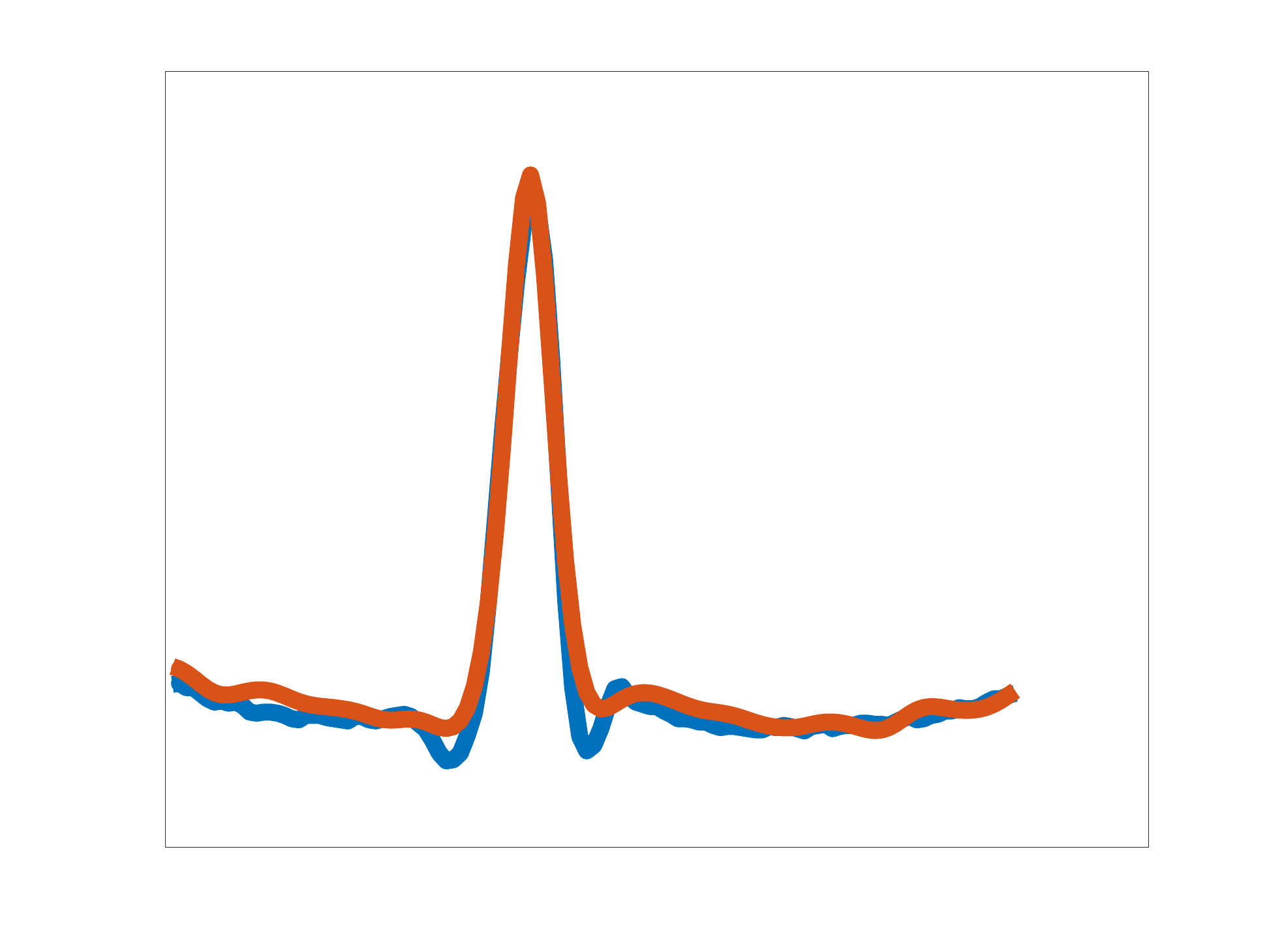}
	\end{overpic}};
\draw[color=black] (0.715,0.19) -- (0.845,0.05);
\draw[color=black] (0.75,0.31) -- (0.845,0.18);
\end{scope}
\end{tikzpicture}
    \vspace{-7mm}
    \caption{\textbf{Enhancement for ECG signals}. From top to bottom: \textbf{(a)} Real noise plot, with muscle artifacts (in red) and electrode motion artifacts (in blue). \textbf{(b)} Contaminated ECG signal (SNR = 9.0678 dB). \textbf{(c)} SCSA denoising method (SNR = 11.4315 dB). In the last graphs, the reconstructed signal (in blue) and the original signal (in red) are superimposed for comparison purposes.}\label{fig:ECG_enhance} 
    \vspace{-0.3cm}
\end{figure}
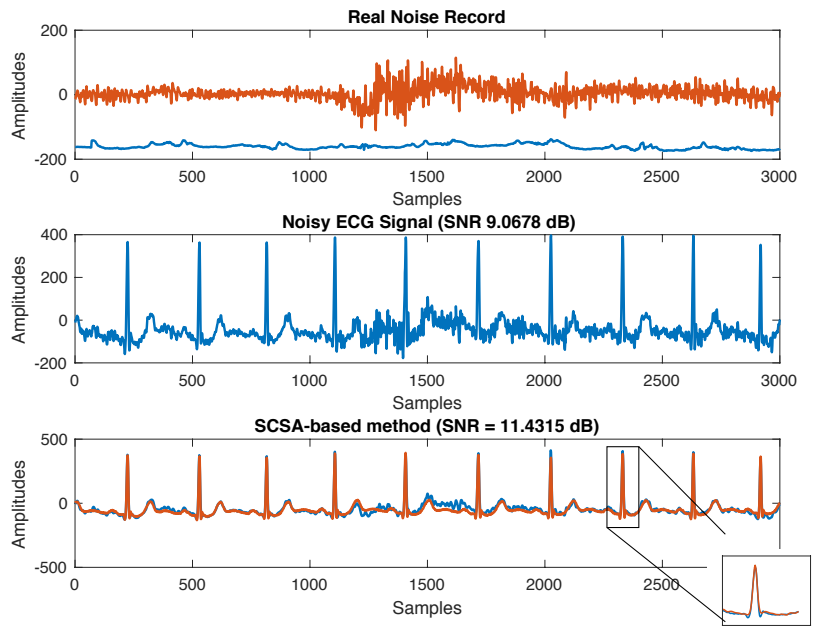

\noindent  In this section, we will consider the ECG case corrupted by
real noise. Real noise records are taken from the MIT-BIH noise stress test database~\cite{b17}. For ECG signal, the analog recordings were played back on a Del Mar Avionics model 660 unit during the digitalization process. The records selected were played back at real time using a specially constructed capstan for the model 660 unit. The analog outputs of the playback unit were filtered to limit analog-to-digital converter (ADC) saturation and for anti-aliasing, using a passband from 0.1 to 100 Hz relative to real time, which is well beyond the lowest and highest frequencies recoverable from the recordings. The bandpass-filtered signals were digitized at 360 Hz per signal relative to real time. Let $n_{ma}(t)$ and $n_{em}(t)$ be the muscle artifact record and the electrode motion record, respectively. The total noise utilized to corrupt the original clean signal $y(t)$ is obtained as $n(t) = k_1n_{ma}(t) + k_2n_{em}(t)$ ($k_i, i = 1,2,\cdots$). $k_1$ and $k_2$ are combined in a way that reaches the same initial SNR in Table~\ref{table}, with a case example shown in Fig.~\ref{fig:ECG_enhance}.

\noindent  Finally, the denoising test is repeated under the same circumstances with different records at different SNRs. The results are presented in Table~\ref{table} in terms of SNR after denoising with corresponding methods, with similar way of selecting parameters as in the simulation part. As can be observed here, the Savitsky-Golay method shows less ability to deal with real noise denoising than the SCSA and wavelet-based method. 

\begin{figure*} [htbp]
    \centering
  \subfloat[Doppler Signal\label{signal}]{%
       \includegraphics[width=0.5\linewidth]{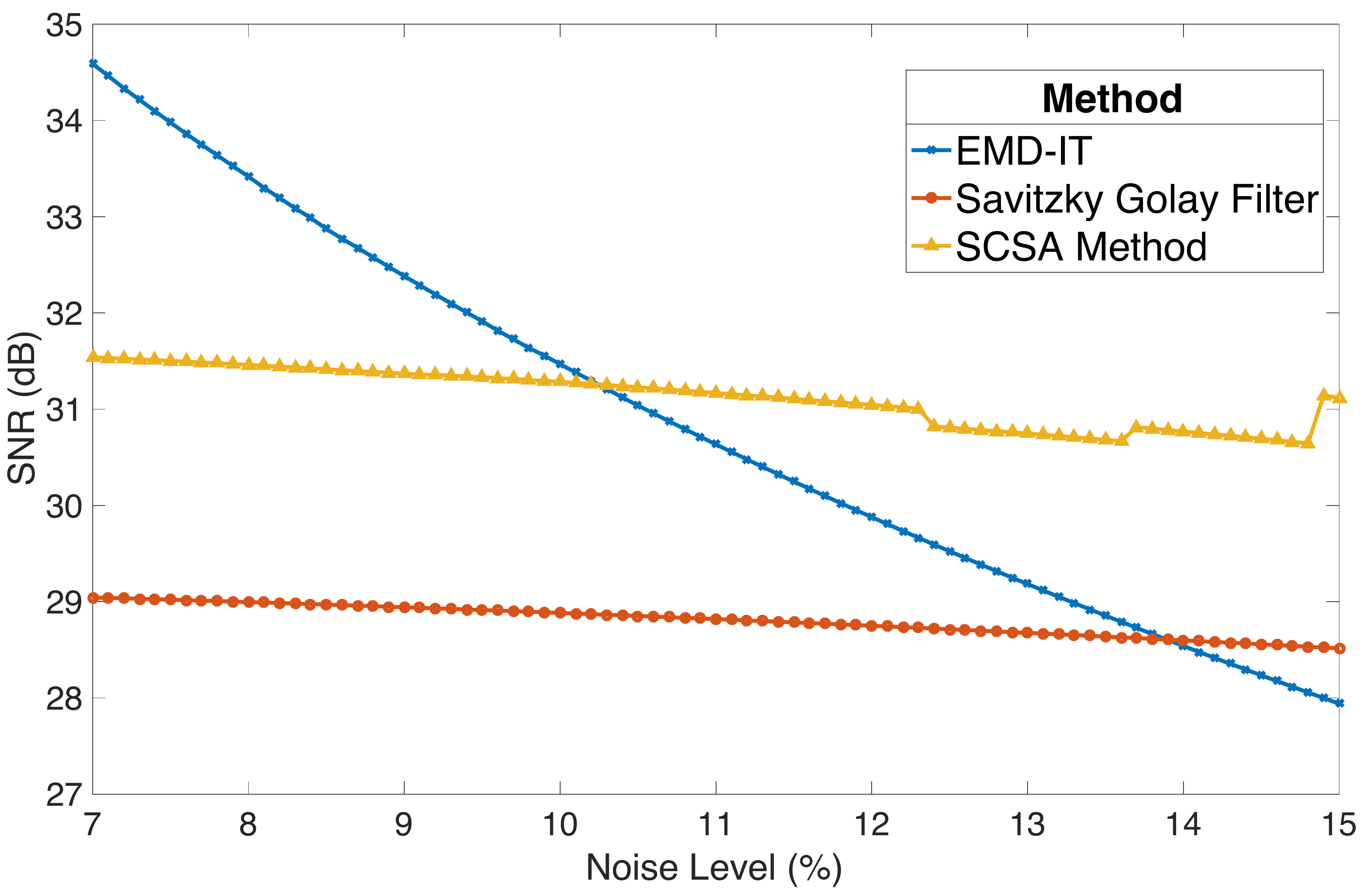}}
    \hfill
  \subfloat[Piecewise-Regular Signal\label{eigenfunction}]{%
        \includegraphics[width=0.5\linewidth]{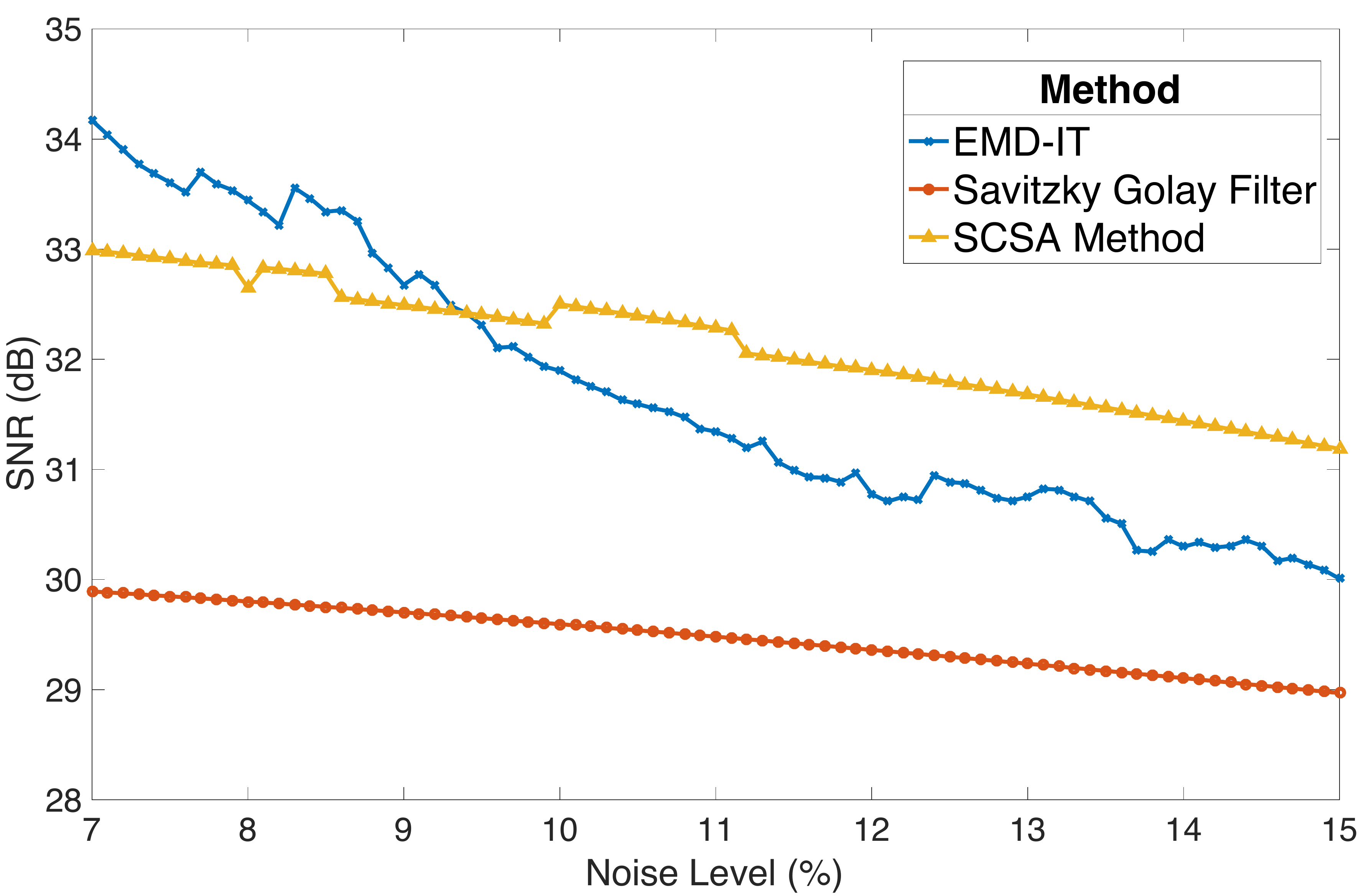}}
            \hfill
  \subfloat[Blocks Signal\label{eigenfunction}]{%
        \includegraphics[width=0.5\linewidth]{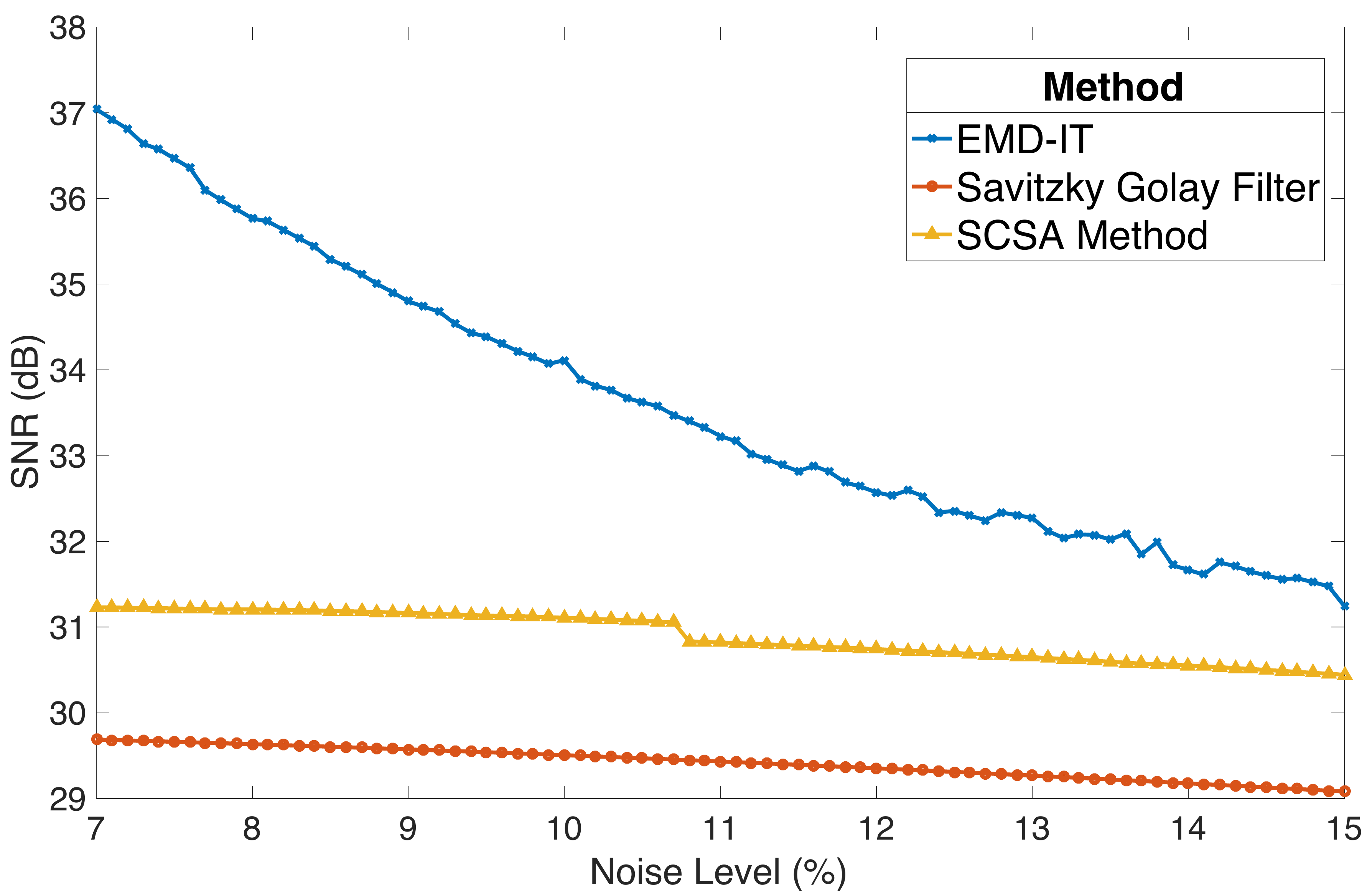}}
            \hfill
  \subfloat[Sing Signal\label{eigenfunction}]{%
        \includegraphics[width=0.5\linewidth]{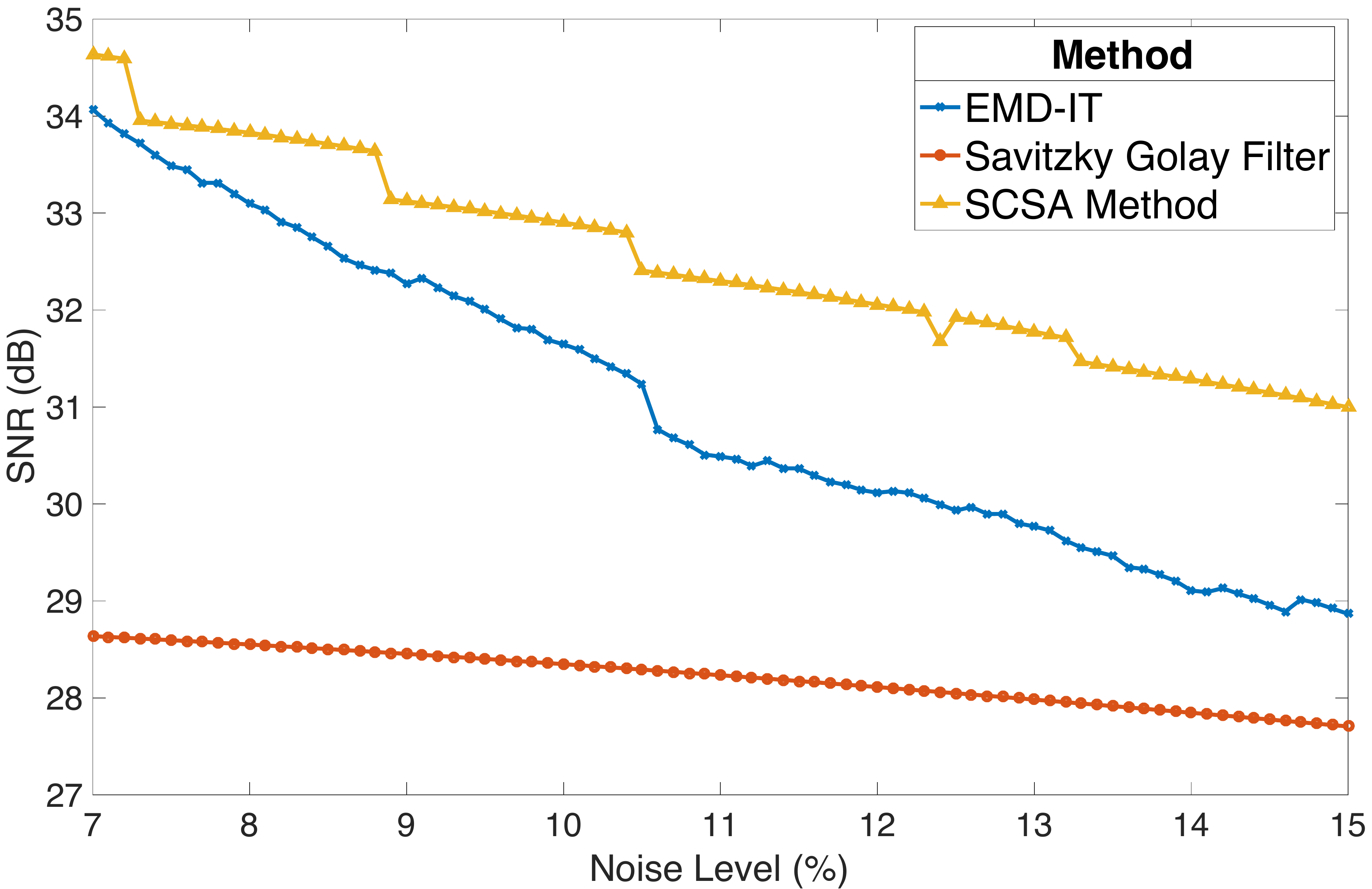}}

  \caption{\textbf{Denoising performance for different types of signals }. \textbf{(a)(b)(c)(d)} SCSA, EMD-Interval thresholding and Savitzky Golay filter denoising performance under different noise level (Noise level ranges between 7\%-15\% (with interval 0.1\%) as shown in the horizontal axis).}
  \label{fig:1234} 
\end{figure*}
\begin{figure*} [htbp]
\vspace{-6mm}
    \centering
  \subfloat [Bumps Signal\label{signal}]{%
       \includegraphics[width=0.5\linewidth]{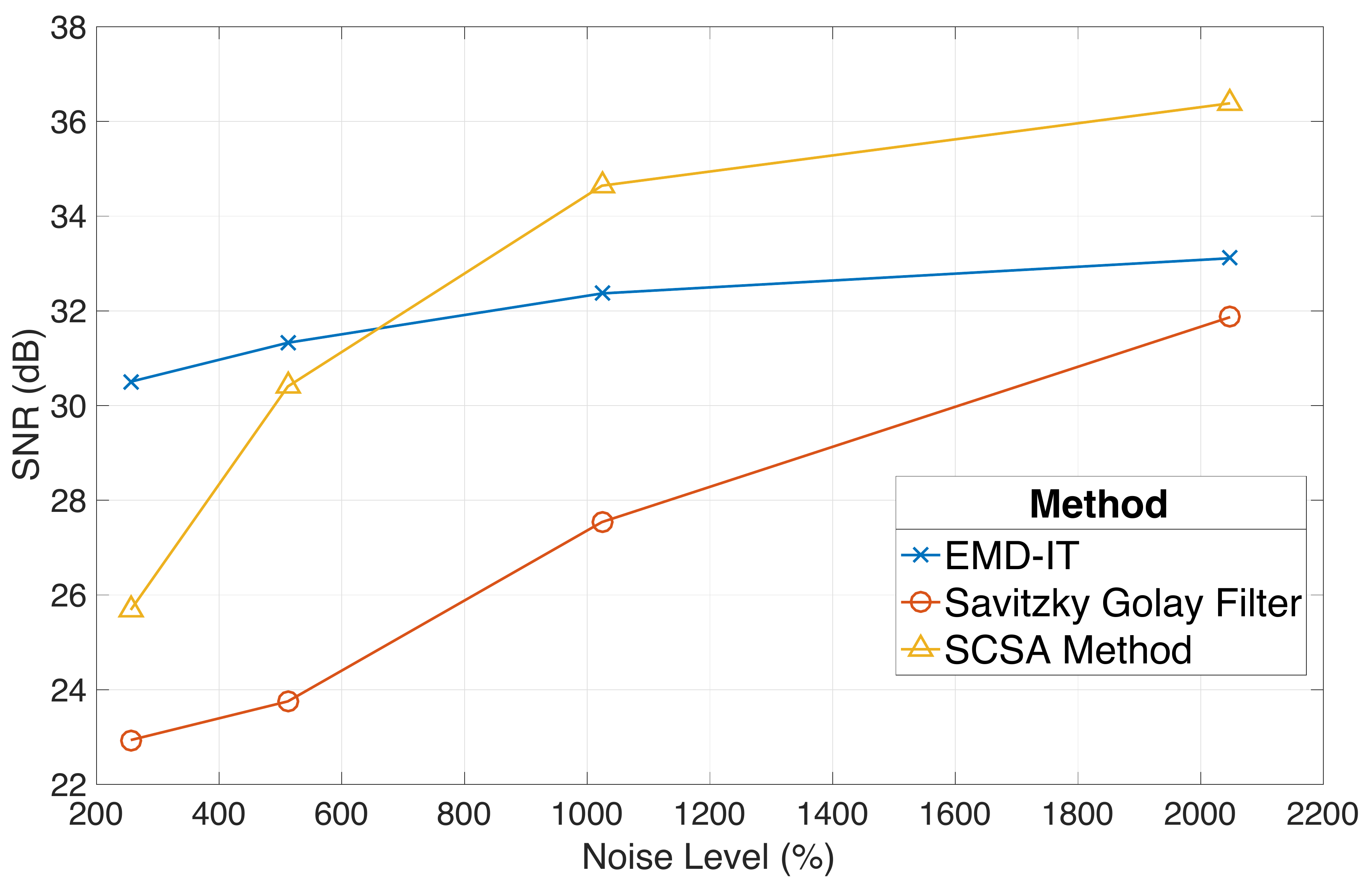}}
    \hfill
  \subfloat[Piecewise-Regular Signal\label{eigenfunction}]{%
        \includegraphics[width=0.5\linewidth]{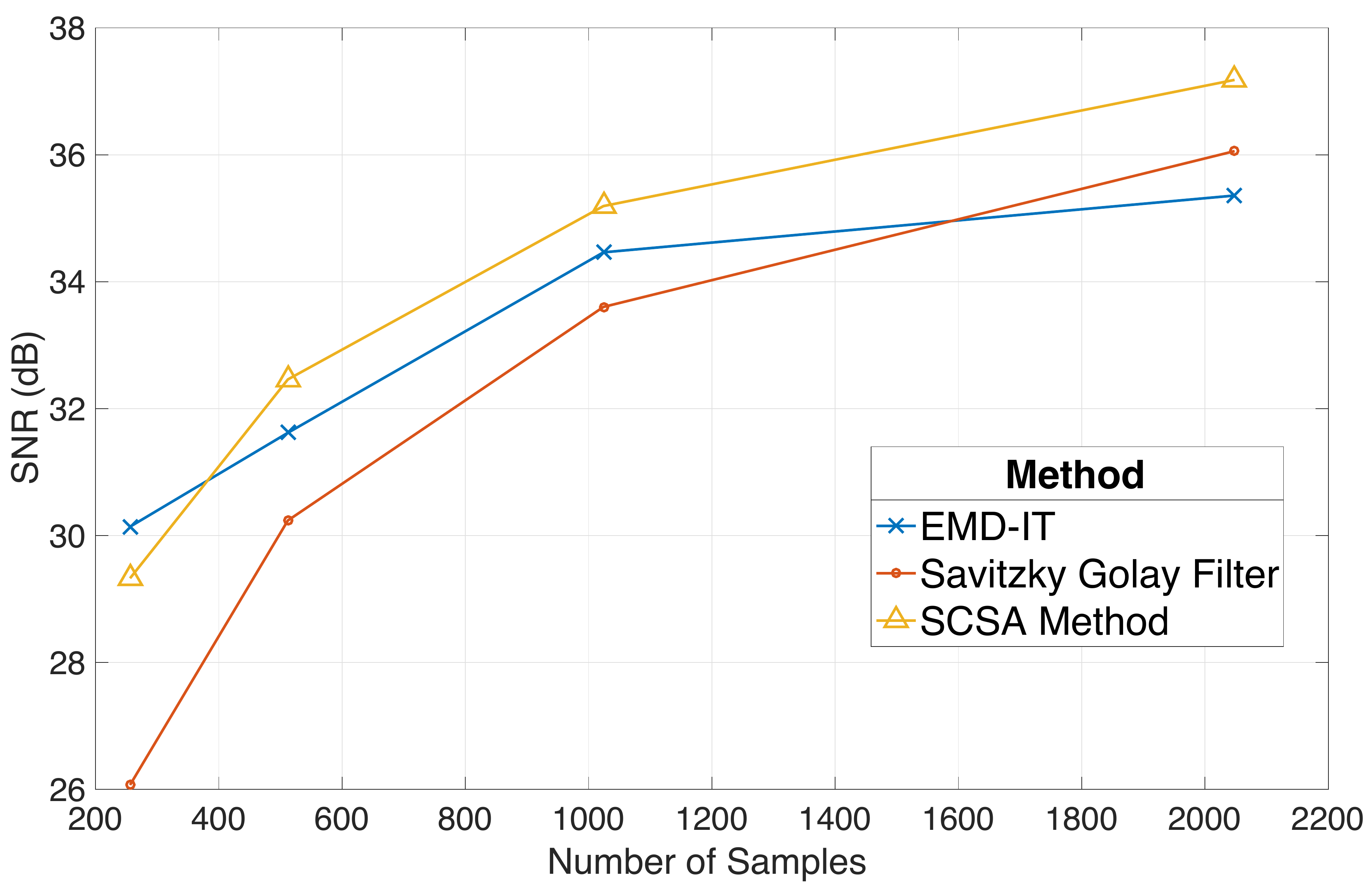}}
            \hfill
  \subfloat[Blocks Signal\label{eigenfunction}]{%
        \includegraphics[ width=0.5\linewidth]{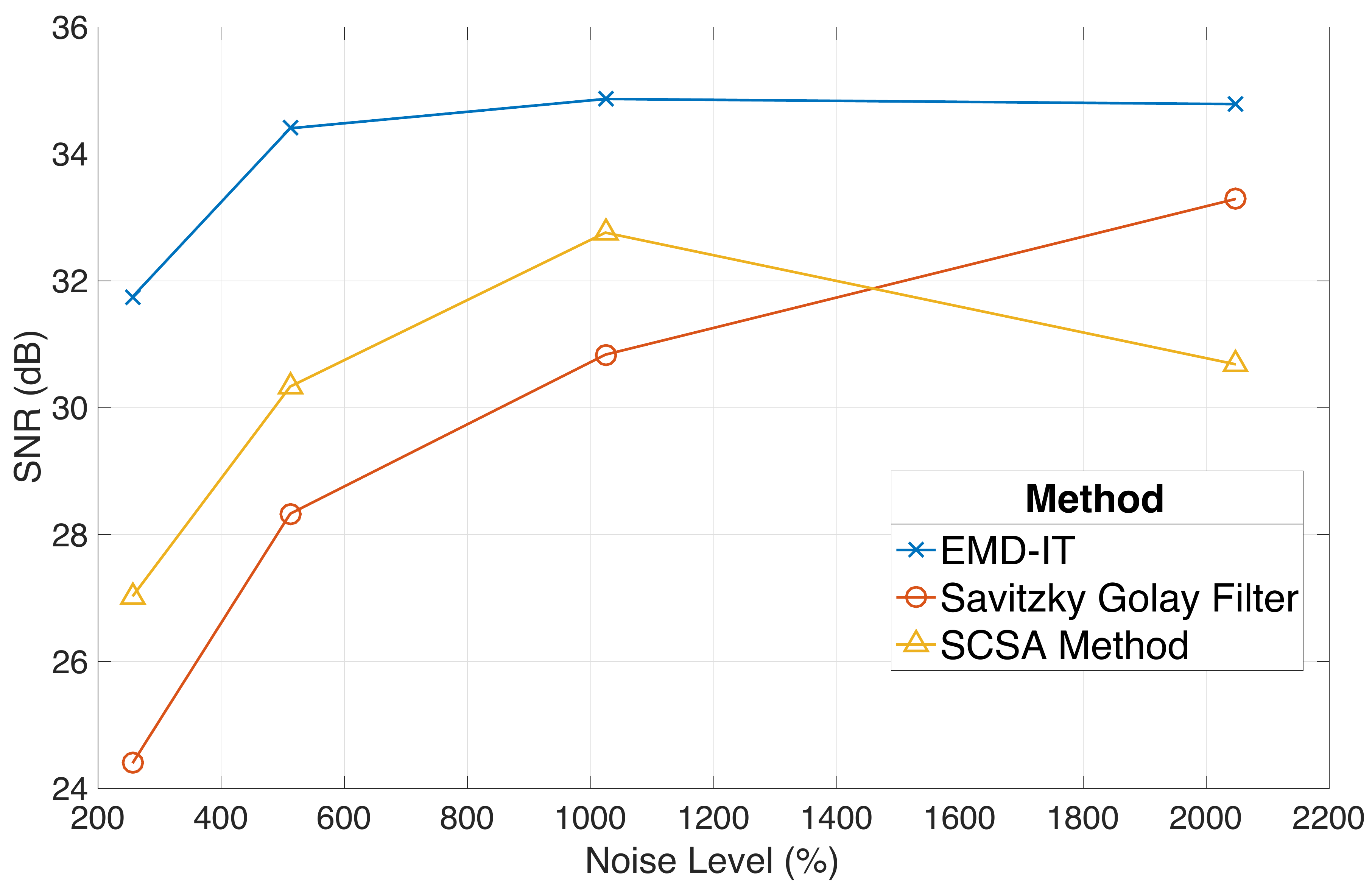}}
            \hfill
  \subfloat[Sing Signal \label{eigenfunction}]{%
        \includegraphics[width=0.5\linewidth]{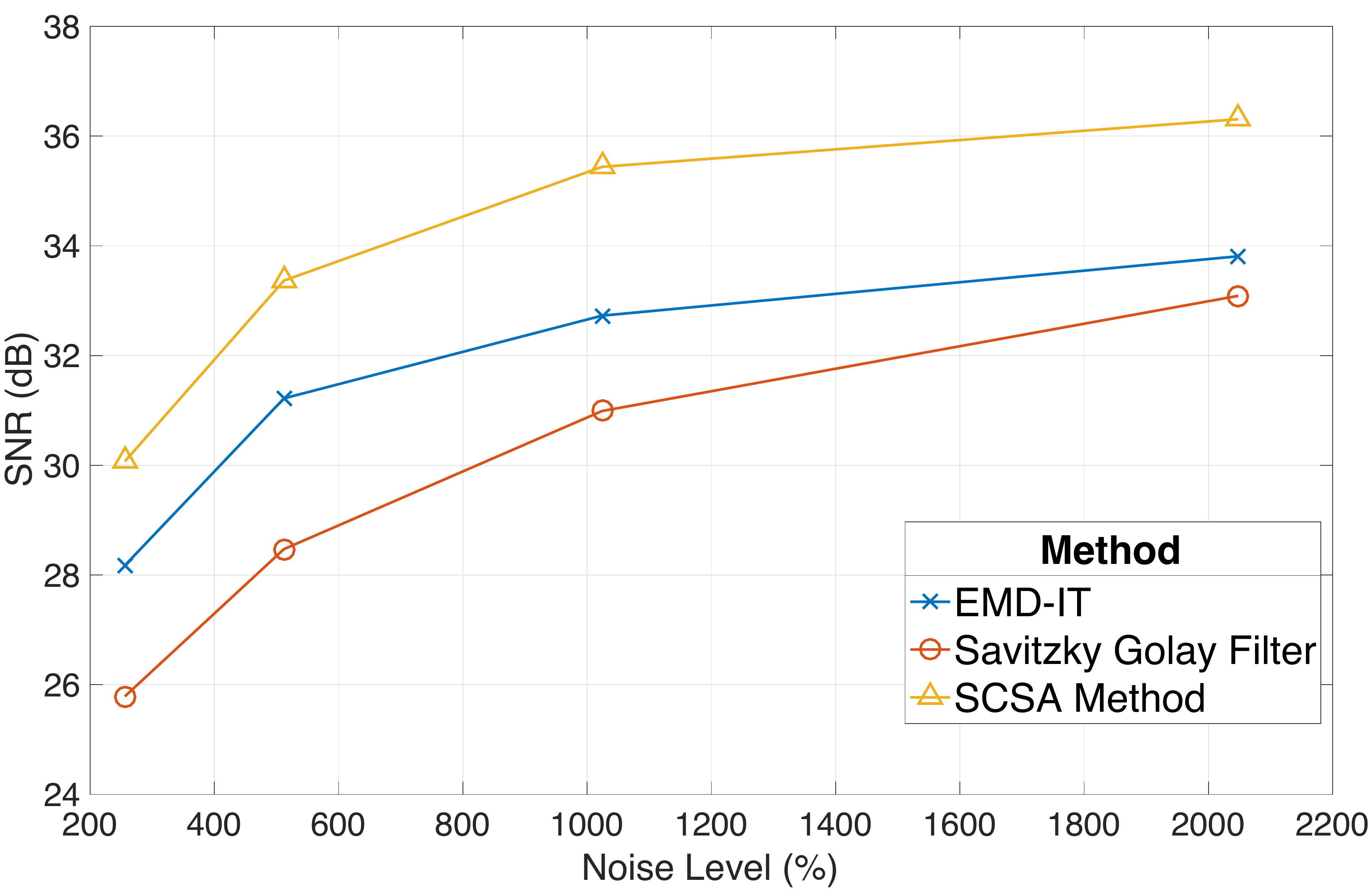}}

  \caption{\textbf{Denoising performance for different sampling frequencies }. \textbf{(a)(b)(c)(d)} SCSA, EMD-Interval thresholding and Savitzky Golay filter denoising performance under 10\% noise level.}
  \label{fig:1234} 
  \vspace{-2mm}
\end{figure*} 

\subsection{General signal denoising}
\begin{figure}[htbp]
  \centering
\includegraphics[width=0.8\textwidth]{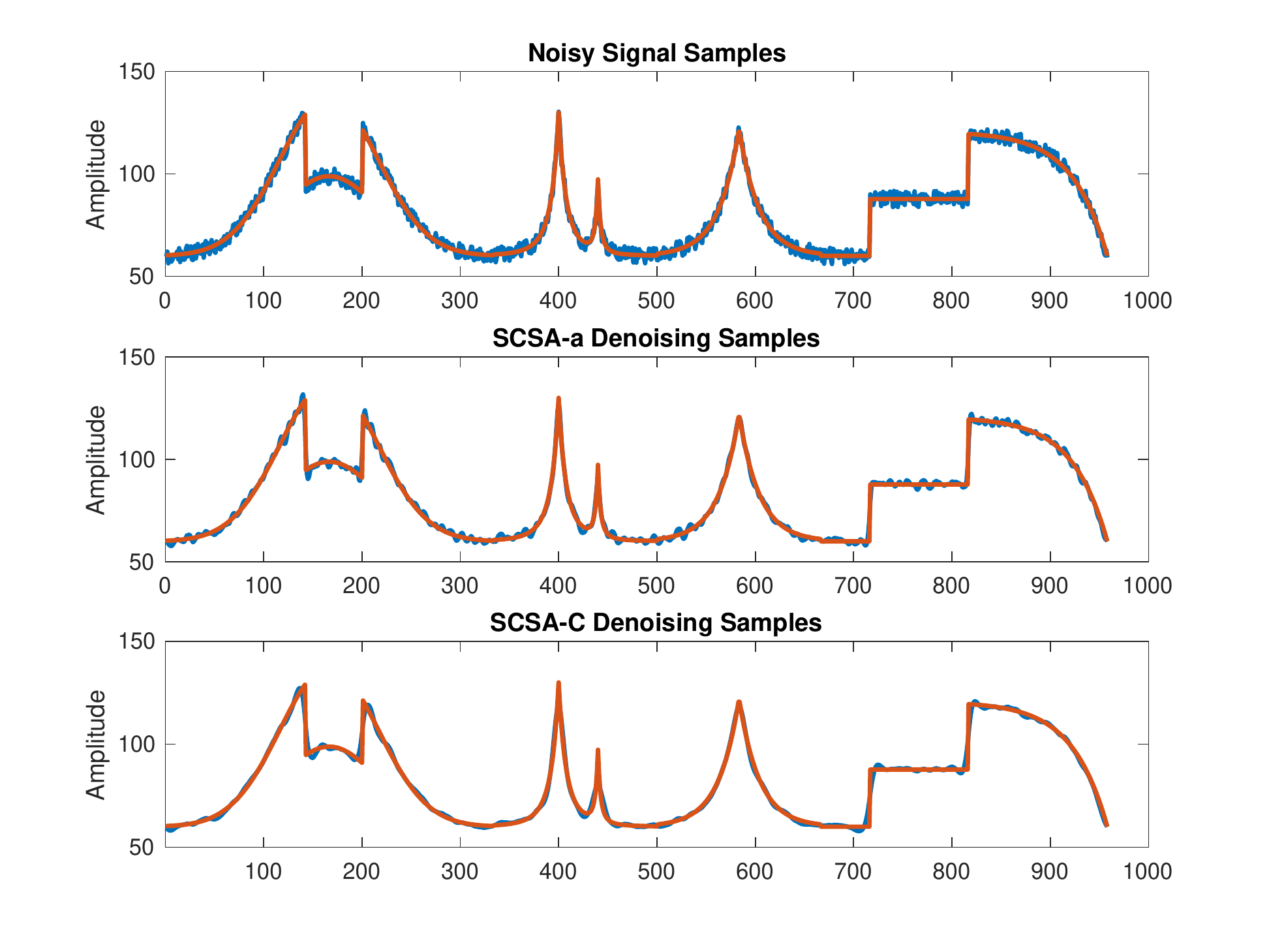}
 \vspace{-0.75cm}
    \caption{\textbf{$\alpha$-SCSA and C-SCSA denoising implementation of the piecewise-regular signal}. From top to bottom: \textbf{(a)} Clean signal (in red) and noisy signal (in blue). \textbf{(b)} Results of the C-SCSA denoising with clean signal (in red) and denoised signal (in blue) (SNR = 30.2657 dB). \textbf{(c)} Results of the C-SCSA denoising with clean signal (in red) and denoised signal (in blue) (SNR = 34.2457 dB). In \textbf{(b)}, $[500,667]$ are identified by the algorithm as peak regions.}\label{fig:SCSA_enhance} 
    \vspace{-0.3cm}
\end{figure}
\begin{figure}[htbp]

  \centering
\includegraphics[width=0.8\textwidth]{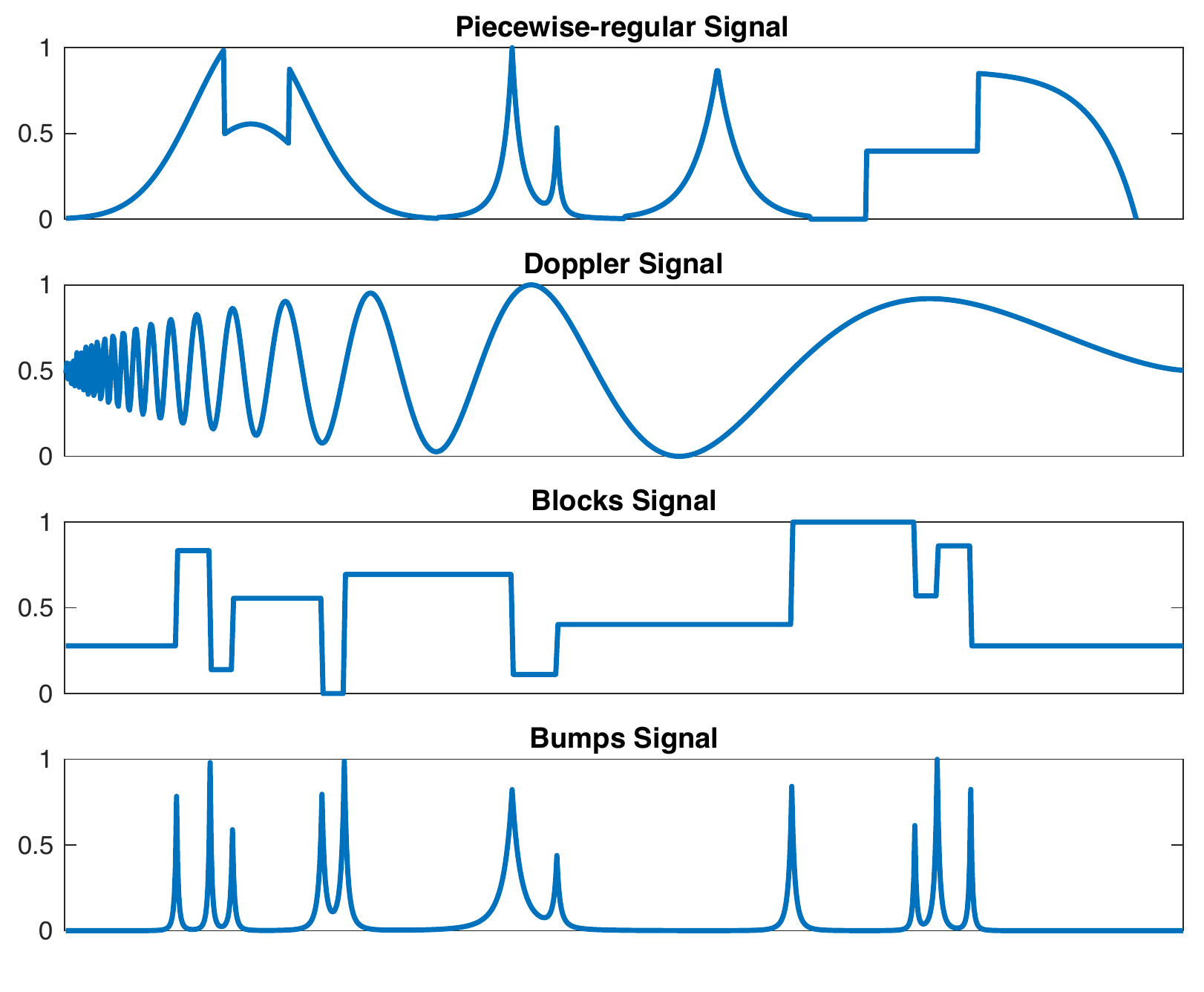}

    \caption{Signals used for validation of different denoising methods. }\label{fig:signal_enhance} 
    \vspace{-0.3cm}
\end{figure}

\noindent Apart from the piecewise-regular signal, three more representative test signals shown in Fig.~\ref{fig:signal_enhance} have been used for validation of the SCSA denoising techniques and others. 

\noindent While it is proved that C-SCSA is also effective in denoising pulse-shaped signals compared to other popular methods, the  C-SCSA real contribution lies in the fact that it can be applied to more general types of signals. Fig.~\ref{fig:SCSA_enhance} depicts an example of the well studied piecewise-regular signal corrupted by white Gaussian noise corresponding to 31.2446 dB signal-to-noise power ratio (SNR). As can be seen, while C-SCSA provides an enhancement of the noisy signal, $\alpha$-SCSA failed in denoising in this case.

\noindent To start with, the effect on the denoising performance of different types of signals with different methods are shown in Fig.~\ref{fig:1234}. We choose the traditional EMD thresholding when comparing to SCSA. Both methods'  parameters are carefully tuned during the simulation, i.e. an optimal universal parameter set is selected for each specific types of signal. More specifically, the adopted performance measure is the SNR after denoising when the noise level before denoising is ranging from  0 \% to 15\% [Fig.~\ref{fig:1234}(a) and (d)]  and the signals used are the Doppler, piecewise regular, blocks and the sing signal [Fig. 8(a)(b)(c)(d)], all sampled with sampling frequency that results in 500 samples. The results shown correspond to ensemble average of 90 independent noise generalizations. The blue curves correspond to the EMD-IT method, the yellow curves to EMD-CIIT and the red curves to traditional SG filter. The triangles ,squares and circles correspond to different simulation sets, respectively. A number of conclusions can be drawn. First, when the signal is regular and vary more slowly, such as the piecewise-regular signal or the sing signal, the SCSA method is performing better than EMD-IT method by showing larger SNR after denoising  and SCSA is better at discarding the noise when the noise level is high. In contrast, when the signal has irregularities, e.g., the blocks signal and the Doppler signal, the best denoising performance (especially in the blocks signal case) is achieved with EMD denoising method (with SCSA method doing better when the noise level increasing to over 10\% in the piecewise-regular signal case). These results have been evaluated with other regular and irregular signals. In general, a more competitive denoising performance is shown by SCSA denoising method when the signal tends to be more regular and slow varying. Secondly, it is apparent that different noise levels do not have significant effect over SCSA method in the Doppler signal case,  since the denoising performance (which is shown in SNR after denoising) difference never exceeds 1 dB. Therefore. there are certain cases that the SCSA method can denoising the signal to a level regardless of the initial noise levels, which is another important characteristics of this method. In addition, the traditional digital filters (shown by SG filter) seems to have less denoising performance compared to the SCSA method and EMD-IT. This happens because, in this case when signal tend to be fast varying,  as shown in Fig.~\ref{fig:signal_enhance}, a tiny shift cased by the filter will damage the SNR through convolution process, while SCSA can preserve peak information of the signal better, as shown in the fourth section of this paper.

\noindent For the rest of simulation examples, each one of the artificial test signals is sampled and tested with four different sampling frequencies to generate four versions per signal, having 256, 512, 1024 and 2048 samples. As before, the results shown correspond to white Gaussian noise generalizations, and in all SCSA-based denoising methods, the penalty parameter is naturally set to make sure that fidelity term and penalty term are of the same order of magnitude.  The adopted performance measure is the SNR after denoising, which corresponds to noise levels of 10\% before denoising. The performance results for different methods shown correspond to Savitzky-Golay and EMD adaptive interval  thresholding. The conclusions drawn from the results are that SCSA-C provide better denoising performance in most cases regardless of the changing of sampling frequencies.

\section{Conclusion}
\noindent The SCSA method combined with a curvature constraint is proposed as a general method to accomplish peak-preserving smoothing task. Details of the SCSA denoising algorithm and its implementation are given in this work.  By designing a proper cost function, we can use the SCSA method to reduce noise while preserving peaks shape. The performance of the proposed method has been investigated and comparison with state of the art methods has been provided.   The method not only produces good denoising performance but also guarantees the peaks are well preserved in the denoising process.

\section*{Acknowledgement}
The research reported in this publication was supported by King Abdullah University of Science and Technology (KAUST) Base Research Fund, (BAS/1/1627-01-01).

 \section{Appendix}
\noindent Let's define $(y_1, y_2, \cdots, y_n)$ to be input signal $\textbf{y}$ with noise. Let $x=y_{m+1}-y_m$ and $w=y_m-y_{m-1}$. Since $x$ and $w$ are subtraction between two neighborhood samples, they resemble noise distributions in homogeneous regions of the signal. Without loss of generality, let's assume $x$ and $w$ to be jointly Gaussian and zero mean with same variance $\sigma^2$, with their joint distribution defined as:
\begin{equation}
f(x,w) = \frac{\exp(-\frac{x^2+w^2-2\rho xw}{2\sigma^2(1-\rho^2)})}{2\pi\sigma^2\sqrt{1-\rho^2}}
\end{equation}
where $\rho = \frac{COV\{x,w\}}{\sigma^2}$. If the signal are equally spaced with interval denoted as $\Delta$, the curvature defined in (4) can be approximated as:
\begin{equation}
k = \frac{|x-w|}{\Delta^2(1+\frac{(x+w)^2}{4\Delta^2})^{\frac{3}{2}}}
\end{equation}
Let $C=\frac{1}{2\pi\sigma^2\sqrt{1-\rho^2}}$, we then give the mathematic induction of $E\{k\}$ as:
\small{\begin{align} \label{eq1}
E\{k\} 
& = \int_{-\infty}^{\infty}\int_{-\infty}^{\infty} \!  k f(x,w) \, \dd x \dd w \nonumber\\
& = \frac{C}{\Delta^2}\int_{-\infty}^{\infty}\int_{-\infty}^{\infty} \!  \frac{|x-w|}{(1+\frac{(x+w)^2}{4\Delta^2})^{\frac{3}{2}}}
\exp\left(-\frac{x^2+w^2-2\rho xw}{2\sigma^2(1-\rho^2)}\right) \dd x \dd w
\end{align}}
Let $\alpha=x+w$, $\beta=x-w$, then $\dd \alpha \dd \beta = 2 \dd x \dd w$. Hence:
\begin{align}\label{equ:E_ab}
&E\{k\}\nonumber\\
& = \frac{C}{2\Delta^2}\int_{-\infty}^{\infty}\int_{-\infty}^{\infty} \! \frac{|\beta|}{(1+\frac{\alpha^2}{4\Delta^2})^{\frac{3}{2}}}
\exp\left(-\frac{(1-\rho)\alpha^2+(1+\rho)\beta^2}{4\sigma^2(1-\rho^2)}\right) \, \dd \alpha \dd \beta \nonumber\\
& = \frac{C}{\Delta^2}
\int_{-\infty}^{\infty} \! \frac{1}{(1+\frac{\alpha^2}{4\Delta^2})^{\frac{3}{2}}}
\exp\left(\frac{-\alpha^2}{4\sigma^2(1+\rho)}\right) \, \dd \alpha 
\int_{0}^{\infty} \! 
\beta \exp\left(\frac{-\beta^2}{4\sigma^2(1-\rho)}\right) \, \dd \beta
\end{align}
Since 
\begin{eqnarray}
\int_{0}^{\infty} \! \beta \exp\left(\frac{-\beta^2}{4\sigma^2(1-\rho)}\right) \, \dd \beta
&=& -2\sigma^2(1-\rho) \left. \exp\left(\frac{-\beta^2}{4\sigma^2(1-\rho)}\right) \right|_{0}^{\infty}\nonumber\\
&=& 2\sigma^2(1-\rho)
\end{eqnarray}

Then Eq.~\eqref{equ:E_ab} reduces to:
\begin{align}
E\{k\} & = \frac{2}{\pi \Delta^2}\sqrt{\frac{1-\rho}{1+\rho}}
\int_{0}^{\infty} \! \frac{1}{(1+\frac{\alpha^2}{4\Delta^2})^{\frac{3}{2}}}
\exp\left(\frac{-\alpha^2}{4\sigma^2(1+\rho)}\right) \, \dd \alpha \nonumber \\
& = \frac{4}{\pi \Delta}\sqrt{\frac{1-\rho}{1+\rho}}
\int_{0}^{\infty} \! \frac{1}{(1+\eta^2)^{\frac{3}{2}}}
\exp\left(-\frac{\Delta^2}{\sigma^2(1+\rho)}\eta^2\right) \, \dd \eta
\end{align}
where $\eta = \alpha/(2\Delta)$. 


\noindent
Note that $k>0$ in all cases. Since $E\{k\} = -\frac{4}{\pi \Delta}\sqrt{\frac{1-\rho}{1+\rho}}\int_{0}^{\infty} \! \frac{1}{(1+\eta)^{\frac{3}{2}}}\exp\left(\frac{\Delta^2}{\sigma^2(1+\rho)}\eta^2\right) \, \dd\eta$, one can easily infer that when $\sigma^2$ increases, $E\{k\}$ also increases. Therefore, in order to reduce noise level $\sigma^2$, we penalize its curvature term $k$.


\begin{thebibliography}{9}
\bibitem{b1} 
T. M. Laleg-Kirati, J. Zhang,  E. Achten, and H. Serrai, "Spectral data de‐noising using semi‐classical signal analysis: application to localized MRS'',    NMR Biomed, vol.29,  pp. 1477-1485, 2016.
\bibitem{b2} 
T. M. Laleg-Kirati, E. Crépeau, M. Sorine, "Semi-classical signal analysis'',  Math. Control Signals Syst, 2013.
\bibitem{b3} 
T. M. Laleg-Kirati, Medigue. C, Cottin F, Sorine. M, "Arterial blood pressure
analysis based on scattering transform II'',  Proceedings of EMBC Sciences
and Technologies for Health, Lyon, France, pp. 3618-3629, 2007.

\bibitem{b4} T. M. Laleg-Kirati, C. Medigue, Y. Papelier, et al, 
"Validation of a semi-classical signal analysis method for stroke volume variation assessment: A comparison with the PiCCO technique
Ann Biomed Eng'',  vol.38,  pp. 3618-3629, 2010. 
\bibitem{b5} A. Chahid, S. Bhaduri, M. Maoui,  R Achten  H. Serrai and T.M. Laleg-Kirati, "Residual Water Suppression Using the Squared
Eigenfunctions of the Schrodinger Operator ", IEEE access, Vol. 9, 69126--69137, 2019. 
\bibitem{b6} 
A. Chahid,T. N. Alotaiby, S. Alshebeili and T-M Laleg-Kirati, "Feature Generation and Dimensionality Reduction using the Discrete
Spectrum of the Schrödinger Operator for Epileptic Spikes Detection",  In Proceeding IEEE EMBC July 2019.  
\bibitem{b7} 
A. F. Kadjo, P. K. Dasgupta, J. Su, S. Y. Liu, and K. G. Kraiczek, "Width Based Quantitation of Chromatographic Peaks: Principles and Principal Characteristics'',  Analytical Chemistry, vol. 89  no.7, pp. 3884-3892, 2017.

\bibitem{b8}

 T. Hastie,  R. Tibshirani, Generalized Additive Models,  Chapman and Hall,  1990.
\bibitem{b9}
M. Bertalmío, S. Levine, "Denoising an image by denoising its curvature image'',  SIAM J. Imag. Sci., vol. 7, no. 2, pp. 187-201, 2014.

\bibitem{b10}
V. Pandey and V. K. Giri, "High frequency noise removal from ECG using moving average filters'', International Conference on Emerging Trends in Electrical Electronics \& Sustainable Energy Systems (ICETEESES), Sultanpur, pp. 191-195, 2016.

\bibitem{b11}
P. Kowalski and R. Smyk, "Review and comparison of smoothing algorithms for one-dimensional data noise reduction'', 2018 International Interdisciplinary PhD Workshop (IIPhDW), Swinoujście, pp. 277-281, 2018.
\bibitem{b12}
Abraham. Savitzky and M. J. E. Golay, "Smoothing and Differentiation of Data by Simplified Least Squares Procedures'',  Analytical Chemistry, vol. 36,  no. 8, pp. 1627-1639,  1964.

\bibitem{b13}
D. Acharya, A. Rani, S. Agarwal, V. Singh,
"Application of adaptive Savitzky–Golay filter for EEG signal processing'',
 Perspectives in Science, vol. 8, pp. 677-679, 2016.
\bibitem{b14}
Y. Li, Y. Ding, T. Li,  "Nonlinear diffusion filtering for peak-preserving smoothing of a spectrum signal'',   Chemometrics and Intelligent Laboratory Systems, vol. 156, pp. 157-165, 2016.

\bibitem{b15}
B. Zimmermann, A. Kohler, "Optimizing Savitzky-Golay parameters for improving spectral resolution and quantification in infrared spectroscopy'', Appl. Spectrosc, vol. 67,  pp. 892-902, 2013.

\bibitem{b16}
B. N. Singh, A. K. Tiwari,
"Optimal selection of wavelet basis function applied to ECG signal denoising'',  Digital Signal Processing, vol. 16, no. 3,  pp.  275-287, 2006.

\bibitem{b17}
A. L. Goldberger, L. A. N. Amaral, L. Glass, J. M. Hausdorff, P. C. Ivanov,
R. G. Mark, J. E. Mietus, G. B. Moody, C. K. Peng, H. E. Stanley,
 "PhysioBank, PhysioToolkit, and PhysioNet: components of a new
research resource for complex physiologic signals",   vol. 101, no. 23, pp. 215-220, 2000.

\bibitem{b18}
Y. Kopsinis and S. McLaughlin, "Development of EMD-Based Denoising Methods Inspired by Wavelet Thresholding,"  IEEE Transactions on Signal Processing, vol. 57, no. 4, pp. 1351-1362, April 2009.




\end{thebibliography}
\end{document}